\documentclass
[twocolumn]
{aa}
\usepackage{epsfig}
\input psfig.sty
\begin{document}

\title{Probing the cosmic distance duality relation with the
       Sunyaev-Zel'dovich effect, X-ray observations and supernovae Ia}
\author{
R.\ F.\ L.\ Holanda\inst{1}\thanks{\email{holanda@astro.iag.usp.br}}
\and
J.\ A.\ S.\ Lima\inst{1}\thanks{\email{limajas@astro.iag.usp.br}}
\and
M.\ B.\ Ribeiro\inst{2}\thanks{\email{mbr@if.ufrj.br}}
      }

\institute{
Departamento de Astronomia, Instituto Astron\^{o}mico e Geof\'{\i}sico,
Universidade de S\~{a}o Paulo -- USP, S\~{a}o Paulo, Brazil
\and
Instituto de F\'{\i}sica, Universidade Federal do Rio de Janeiro -- UFRJ,
Rio de Janeiro, Brazil
         }

%\date{Accepted . Received ; in original form }

\date{}

\abstract {The angular diameter distances toward galaxy clusters
can be determined with measurements of Sunyaev-Zel'dovich effect
and X-ray surface brightness combined with the validity of the
distance-duality relation, $D_L(z) (1 + z)^{2}/D_{A}(z) = 1$,
where $D_L(z)$ and $D_A(z)$ are, respectively, the luminosity and
angular diameter distances. This combination enables us to probe
galaxy cluster physics or even to test the validity of the
distance-duality relation  itself.}{We explore these possibilities
based on two different, but complementary approaches. Firstly, in
order to constrain the possible galaxy cluster morphologies, the
validity of the distance-duality relation (DD relation) is assumed
in the $\Lambda$CDM framework (WMAP7). Secondly, by adopting a
cosmological-model-independent test, we directly confront the angular
diameters from galaxy clusters with two supernovae Ia (SNe Ia)
subsamples (carefully chosen to coincide with the cluster positions).
The influence of the different SNe Ia light-curve fitters in the
previous analysis are also discussed.}{We assumed that $\eta$ is a
function of the redshift parametrized by two different relations:
$\eta(z) = 1 + \eta_{0}z$, and $\eta(z)=1 + \eta_{0}z/(1+z)$, where
$\eta_0$ is a constant parameter quantifying the possible departure
from the strict validity of the DD relation. In order to determine
the probability density function (PDF) of $\eta_{0}$, we considered
the angular diameter distances from galaxy clusters recently studied
by two different groups by assuming elliptical and spherical
isothermal $\beta$ models and spherical non-isothermal $\beta$ model.
The strict validity of the DD relation will occur only if the maximum
value of $\eta_{0}$ PDF is centered on $\eta_{0}=0$.}{For both
approaches we find that the elliptical $\beta$ model agrees with 
the distance-duality relation, whereas the non-isothermal spherical
description is, in the best scenario, only marginally compatible. We
find that the two-light curve fitters (SALT2 and MLCS2K2) present a
statistically significant conflict, and a joint analysis involving
the different approaches suggests that clusters are endowed with an
elliptical geometry as previously assumed}{The statistical analysis
presented here provides new evidence that the true geometry of
clusters is elliptical. In principle, it is remarkable that a local
property such as the geometry of galaxy clusters might be constrained
by a global argument like the one provided by the cosmological
distance-duality relation.}

\authorrunning{R.F.L.\ Holanda et al.\ }
\titlerunning{Probing Cosmic Distance Duality Relation}
\maketitle

\section{Introduction}\label{sec:introduction}

The \textit{reciprocity law} or \textit{reciprocity theorem}, proved
long ago by Etherington (1933), is a fundamental keystone for the
interpretation of astronomical observations in cosmology. It states
that if source and observer are in relative motion, solid angles
subtended between the source and the observer are related by
geometrical invariants where the redshift $z$ measured for the
source by the observer enters in the relation. The core idea of this
law comes from the invariance of various geometrical properties when
there is a transposition between the roles of source and observer in
astronomical observations. Proofs were presented in the context of
relativistic geometrical optics, where it comes as a consequence of
the geodesic deviation equation, as well as in the context of
relativistic kinetic theory, where it is based on the Liouville
integral for collision-free photons. The fundamental hypothesis
behind the reciprocity law is the one made in General Relativity
that light travels along null geodesics in a Riemannian spacetime
(see Ellis 1971, 2007, and references therein).

Etherington's reciprocity law can be presented in various 
ways, either in terms of solid angles or relating various
cosmological distances. Its most useful version in the context of
astronomical observations, sometimes referred to as the
\textit{distance-duality} (DD) relation, relates the
\textit{luminosity distance}, $D_{\scriptstyle L}$, with the
\textit{angular diameter distance}, $D_{\scriptstyle A}$,
by means of the following equation
\begin{equation}
 \frac{D_{\scriptstyle L}}{D_{\scriptstyle A}}{(1+z)}^{-2}=1.
 \label{rec}
\end{equation}
Since this result is easily proved in
Friedmann-Lema\^itre-Robertson-Walker (FLRW) cosmologies, perhaps
this is the reason why the generality of the relation above is not
fully appreciated by most authors. Indeed, this law is
\textit{completely general}, valid for \textit{all} cosmological
models based on Riemannian geometry, being dependent neither on
Einstein field equations nor on the nature of matter. Therefore,
the DD relation is valid for both homogeneous and inhomogeneous
cosmological models, requiring only that source and observer are
connected by null geodesics in a Riemannian spacetime and that the
number of photons is conserved.

The DD relation plays an essential role in modern cosmology, ranging
from gravitational lensing studies (Schneidder, Ehlers \& Falco
1992) to analyses of galaxy distribution and galaxy clusters
observations (Lima, Cunha \& Alcaniz 2003; Cunha, Marassi \& Lima
2007; Rangel Lemos \& Ribeiro 2008; Ribeiro 1992, 1993, 2005;
Ribeiro \& Stoeger 2003; Albani et al.\ 2007; Mantz et al.\ 2010;
Komatsu et al.\ 2011), as well as the plethora of cosmic
consequences from primary and secondary temperature anisotropies of
the cosmic microwave blackbody radiation (CMBR) observations
(Komatsu et al.\ 2011). Other consequences of Etherington's
reciprocity relation are the temperature shift equation
$T_o=T_e/(1+z)$, where $T_o$ is the observed temperature and $T_e$
is the emitted temperature, a key result for analyzing CMBR
observations, as well as the optical theorem that surface brightness
of an extended source does not depend on the angular diameter
distance of the observer from the source, an important result for
understanding lensing brightness (Ellis 2007). In this connection,
we also observe that any source of attenuation, such as ``gray''
intergalactic dust or exotic photon interaction, contributes to
violate the DD relation because its proof is based on the
conservation of the average number of photons.

Although taken for granted in virtually all analyses in cosmology,
Eq. (\ref{rec}) is in principle testable by means of astronomical
observations (Uzan, Aghanim \& Mellier 2004; Basset \& Kunz 2004;
Holanda, Lima \& Ribeiro 2010; Li, Wu \& Yu 2011, Nair, Jhingham \&
Jain 2011). If one is able to find cosmological sources whose
intrinsic luminosities (standard candles) as well as their
intrinsic sizes (standard rulers) are known, one can determine both
$D_{\scriptstyle L}$ and $D_{\scriptstyle A}$ and, after measuring
the redshifts, test the cosmic version of Etherington's result as
given by the equation above. Note that ideally both quantities must
be measured in a way that does not utilize any relationship coming
from a cosmological model, that is, they must be determined by means
of intrinsic astrophysically measured quantities. Therefore, the
ideal way of observationally testing the reciprocity equation
(\ref{rec}) would require \textit{independent} measurements of
intrinsic luminosities and sizes of sources, that is, independent
determinations of $D_{\scriptstyle A}$ and $D_{\scriptstyle L}$ for
a given set of sources.

There are, however, less-than-ideal methods to test Eq. (\ref{rec}).
These usually assume a cosmological model suggested by a set of
observations, apply this model in the context of some astrophysical
effect, thereby trying to see if the DD relation remains valid. In
this context, Uzan, Aghanim \& Mellier (2004) argued that the
Sunyaev-Zel'dovich effect plus X-ray techniques for measuring
$D_{\scriptstyle A}(z)$ from galaxy clusters (Sunyaev \& Zel'dovich
1972; Cavaliere \& Fusco-Fermiano 1978) is strongly dependent on the
validity of this relation (see details in the next section).
Briefly, in the context of this phenomenon one may consider the
different electronic density dependencies combined with some
assumptions about the galaxy cluster morphology in order to evaluate
its angular diameter distance with basis on Eq.\ (\ref{rec}), such
that
\begin{equation}
D_{A}(z)\propto \frac{{D_A}^2(\Delta
T_{0})^{2}\Lambda_{eH0}}{{D_L}^{2}
S_{X0}{T_{e0}}^{2}}\frac{1}{\theta_{c}}\propto \frac{(\Delta
T_{0})^{2}\Lambda_{eH0}}{(1+z)^4
S_{X0}{T_{e0}}^{2}}\frac{1}{\theta_{c}},
\end{equation}
where $S_{X0}$ is the central X-ray surface brightness, $T_{e0}$ is
the central temperature of the intra-cluster medium, $\Lambda_{eH0}$
is the central X-ray cooling function of the intra-cluster medium,
$\Delta T_0$ is the central decrement temperature, and $\theta_{c}$
refers to a characteristic scale of the cluster along the line of
sight (l.o.s.), whose exact meaning depends on the assumptions
adopted to describe the galaxy cluster morphology.

On the other hand, in order to test the validity of DD relation it
is convenient to assume a deformed expression:
\begin{equation}
\frac{D_{\scriptstyle L}}{D_{\scriptstyle A}}{(1+z)}^{-2}= \eta (z),
\label{receta}
\end{equation}
and from Eq.\ (2) we have {(see section 4 for details)}
\begin{equation}
D_{A}^{\: data}(z)=D_{A}(z)\eta(z)^{2}, \label{recc}
\end{equation}
actually, multiplied by $\eta^{-2}$ in the notation of Uzan et al.\
(2004). This quantity is reduced to the standard angular diameter
distance  only when the DD relation is strictly valid ($\eta\equiv
1$). In order to quantify the $\eta$ parameter, Uzan  et al.\ (2004)
fixed $D_A(z)$ by using the cosmic concordance model (Spergel et
al.\ 2003) while for $D^{\: data}_{A}(z)$ they considered the 18
galaxy clusters from the Reese et al.\ (2002) sample for which a
spherically symmetric cluster geometry has been assumed. By assuming
$\eta$ constant, their statistical analysis provided $\eta = 0.91^{+
0.04}_{-0.04}$ (1$\sigma$), and is therefore only marginally
consistent with the standard result.

Basset \& Kunz (2004) used  supernovae Ia (SNe Ia) data as
measurements of the luminosity distance and the estimated
$D_{\scriptstyle A}$ from FRIIb radio galaxies (Daly \& Djorgovski
2003) and ultra compact radio sources (Gurvitz 1994, 1999; Lima \&
Alcaniz 2000, 2002; Santos \& Lima 2008) in order to test possible
new physics  based on the following generalization of Eq.\ (\ref{rec})
\begin{equation}
\frac{D_L(z)}{D_A(z)(1 + z)^{2}} = (1+z)^{\beta-1} \exp\left[\gamma
\int_0^z \frac{dz'}{E(z')(1+z')^{\alpha}}\right], \label{3param}
\end{equation}
where $E(z) \equiv H(z)/H_0$ is the dimensionless Hubble expansion,
a quantity normalized to unity today. Note that for arbitrary
values of $\alpha$, the strict validity of the DD relation
corresponds to $(\beta, \gamma) = (1,0)$.  By marginalizing on
$\Omega_M$, $\Omega_{\Lambda}$ and Hubble parameters, Basset \& Kunz
(2004) found a $2\sigma$ violation caused by excess brightening of
SNIa at $z > 0.5$, perhaps owing to a lensing magnification bias.

On the other hand, De Bernardis, Giusarma \& Melchiorri (2006) also
confronted the angular distances from galaxy clusters with
luminosity distance data from SNe Ia  to obtain a model-independent
test. In order to compare the data sets they considered the weighted
average of the data in seven bins and found that $\eta = 1$ is
consistent on a $68\%$ confidence level (1$\sigma$). However, one
needs to be careful when using the Sunyaev-Zel'dovich effect together
with X-ray techniques for measuring angular diameter distances to
test the DD relation because this technique also depends on
its validity. Indeed, when the relation does not hold, $D_A(z)$
determined from observations is in general $D^{\:
data}_{A}(z)=D_{A}(z)\eta^{2}$, which is reduced to $D_{A}$ only if
$\eta =1$. This means that De Bernardis and co-workers did not really
test the DD relation, at least not in a consistent way. In addition,
these authors binned their data, and, as such, their results may have
been influenced by the particular choice of redshift binning.

Avgoustidis et al.\ (2010) also adopted an extended DD relation,
$D_L=D_A(1+z)^{2+\epsilon}$, in the context of a flat $\Lambda$CDM
model for constraining the cosmic opacity. The recent SN type Ia
data compilation (Kowalski et al.\ 2008) was combined with the
latest measurements of the Hubble expansion at redshifts in the
range $0<z<2$ (Stern et al.\ 2010). They found
$\epsilon=-0.04_{-0.07}^{+0.08}$ (2$\sigma$). It should be stressed,
however, that the main goal of the quoted studies was to merely test
the consistency between the assumed cosmological model and the
results provided by a chosen set of astrophysical phenomena. 

Below we explore a different route to test the DD relation
by using two complementary, but independent, approaches. First, we
take the DD's validity for granted in order to access the galaxy cluster
morphology. The basic idea is very simple and may be described as
follows. The usually assumed spherical geometry of clusters has been
severely questioned after some analyses based on data from the
XMM-Newton and Chandra satellites, which suggested that clusters
are supposed to exhibit an elliptical surface brightness. In this
context, by assuming the $\Lambda$CDM framework (WMAP7), we discuss
the constraints coming from the validity of DD relation on the local
physics, that is, when different assumptions about the cluster
geometry are considered. In order to do that, we considered three
samples of angular diameter
distances from galaxy clusters obtained through Sunyaev-Zel'dovich
effect and X-ray measurements. These samples differ by the
assumptions adopted to describe the clusters: (i) isothermal
elliptical, (ii) isothermal spherical $\beta$ models (De Filippis et
al.\ 2005),  and non-isothermal spherical double $\beta$ model
(Bonamente et al.\ 2006). Secondly, we propose a consistent
cosmological-model-independent test for the DD relation by using
subsamples of SNe Ia carefully chosen from Constitution data
(Hicken et al.\ 2009) and the angular diameter distances from galaxy
clusters. These topics were partially discussed by us (Holanda, Lima
\& Ribeiro 2010, 2011)  without considering the second possibility
(isothermal spherical $\beta$ model) that has also been analyzed by
De Filippis et al.\ (2005). Both approaches were separately
investigated, however, by avoiding all details of the Sunyaev-Zel'dovich
effect and X-ray cluster physics.

In this article,  we intend to close the above described gaps by
providing a closer discussion of the physics involved, and, for
completeness, we also included the spherical $\beta$ model case.
In addition, the influence of the different SNe Ia
light-curve fitters on the model-independent test involving
the Sunyaev-Zel'dovich effect, X-ray and SNe Ia  is discussed.
Finally, a joint analysis involving the different approaches is
also investigated. The present study (based on complementary tests)
suggests that clusters are endowed with an elliptical geometry as
assumed by De Filippis et al.\ (2005), once the strict validity of
DD relation is taken for granted.

The paper is organized as follows. Sunyaev-Zel'dovich effect and
X-ray surface brightness observations as a test for the DD relation
are explored in \S\ref{sec:sze}. The galaxy cluster samples used in
this paper are presented in \S\ref{sec:sample}. The consistence
between the validity of the DD relation and the different
assumptions used to describe the galaxy clusters usually adopted in
the literature are discussed in \S\ref{sec:meth1}. In
\S\ref{sec:meth2} we discuss a new and model-independent
cosmological test for the DD relation involving luminosity distances
from SNe Ia and $D_A(z)$ from galaxy clusters. {In \S 6 we study the
influence of the different SNe Ia light-curve fitters on the results
of the previous section, while the joint analysis is presented in \S
7. Finally, the main conclusions and future prospects are summarized
in \S\ref{sec:conclusions}.

\section{SZE/X-ray technique and the distance-duality relation}
\label{sec:sze}

An important phenomenon occurring in galaxy clusters is the
Sunyaev-Zel'dovich effect (SZE), a small distortion of the CMBR
spectrum provoked by the inverse Compton scattering of the CMBR
photons passing through a population of hot electrons. The SZE is
proportional to the electron pressure integrated along the l.o.s.,
i.e., to the first power of the plasma density. The measured
temperature decrement $\Delta T_{\rm SZ}$ of the CMBR is given by
(De Filippis et al.\ 2005)
\begin{equation}
\label{eq:sze1} \frac{\Delta T_{\rm \scriptscriptstyle SZ}}{T_{\rm
\scriptscriptstyle CMBR}} = f(\nu, T_{\rm e}) \frac{ \sigma_{\rm T}
k_{\rm B} }{m_{\rm e} c^2} \int_{\rm l.o.s.}n_e T_{\rm e} dl, \
\end{equation}
where $T_{\rm e}$ is the temperature of the intra-cluster medium,
$k_{\rm B}$ the Boltzmann constant, $T_{\rm \scriptscriptstyle CMBR}
 =2.728^{\circ}$K
is the temperature of the CMBR, $\sigma_{\rm T}$ the Thompson cross
section, $m_{\rm e}$ the electron mass and $f(\nu, T_{\rm e})$
accounts for frequency shift and relativistic corrections (Itoh,
Kohyama \& Nozawa 1998; Nozawa, Itoh \& Kohyama 1998).

Other important physical phenomena occurring in the intra-galaxy
cluster medium are the X-ray emission caused by thermal bremsstrahlung
and line radiation resulting from electron-ion collisions. The X-ray
surface brightness $S_X$ is proportional to the integral along the
line of sight of the square of the electron density. This quantity
may be written as follows
\begin{equation}
S_X = \frac{D_A^2}{4 \pi D_L^2 } \int _{\rm l.o.s.} n_e^2
\Lambda_{eH} dl , \label{eq:sxb0}
\end{equation}
where $\Lambda_{eH}$ is the X-ray cooling function of the intra-cluster
medium (measured in the cluster rest frame) and $n_e$ is the
electron number density. It thus follows that the SZE and X-ray
emission both depend on the properties ($n_e$, $T_e$) of the intra
cluster medium.

As is well known, it is possible to obtain the angular diameter
distance from galaxy clusters by their SZE and X-ray surface
brightness observations. The calculation begins by constructing a
model for the cluster gas distribution. Assuming, for instance, the
spherical isothermal $\beta$-model such that $n_e$ is given by
(Cavaliere \& Fusco-Fermiano 1978)
\begin{equation}
n_e({{r}}) =  \left ( 1 + \frac{r^2}{r_c^2} \right )^{-3\beta/2},
\label{eq:iso_beta}
\end{equation}
equations (6) and (7) can be integrated. Here $r_c$ is the core
radius of the galaxy cluster. This $\beta$ model is based on the
hydrostatic equilibrium equation and constant temperature (Sarazin
1988). In this way, we may write for the SZE 
\begin{equation}
\label{eq:sz2} \Delta T_{\rm SZ} = \Delta T_0 \left( 1+
\frac{\theta^2 }{\theta_{c}^2} \right)^{1/2-3\beta/2},
\end{equation}
where { $\theta_{c}=r_c/{{D}}_A$ is the angular core radius and}
$\Delta T_0$ is the central temperature decrement that includes all
physical constants and terms resulting from the line-of-sight
integration. More precisely:
\begin{equation}
\label{eqsze3} \Delta T_0 \equiv T_{\rm \scriptscriptstyle CMBR}
f(\nu, T_{\rm e}) \frac{ \sigma_{\rm T} k_{\rm B} T_{\rm e}}{m_{\rm e}
c^2}n_{e0} \sqrt{\pi} \theta_c D_A g\left(\beta/2\right),
\end{equation}
with
\begin{equation}
g(\alpha)\equiv\frac{\Gamma \left[3\alpha-1/2\right]}{\Gamma \left[3
\alpha\right]}, \label{galfa}
\end{equation}
where $\Gamma(\alpha)$ is the gamma function and the others
constants are the usual physical quantities. For X-ray surface
brightness, we have
\begin{equation}
S_X = S_{X0} \left( 1+ \frac{\theta }{\theta_{c}^2} \right)^{1/2-3
\beta}, \label{eqsxb1}
\end{equation}
where the central surface brightness $S_{X0}$ reads
\begin{equation}
\label{eqsxb2} S_{X0} \equiv \frac{D_A^2 \Lambda_{eH}\
\mu_e/\mu_H}{D_L^2 4 \sqrt{\pi} } n_{e0}^2 \theta_c D_A \ g(\beta).
\end{equation}
Here $\mu$ is the molecular weight given by $\mu_i\equiv
\rho/n_im_p$.

One can solve equations (10) and (13) for the angular diameter
distance by eliminating $n_{e0}$ and taking for granted the validity
of DD relation. However, a more general result appears when the DD
relation is not regarded as being strictly valid. In this case one
obtains
\begin{eqnarray}
\label{eqobl7}
{{D}}_A &= & \left[ \frac{\Delta {T_0}^2}{S_{\rm X0}}
\left( \frac{m_{\rm e} c^2}{k_{\rm B} T_{e0} } \right)^2
\frac{g\left(\beta\right)}{g(\beta/2)^2\ \theta_{\rm c}}
\right] \times \nonumber \\
& & \times \left[ \frac{\Lambda_{eH0}\ \mu_e/\mu_H}{4 \pi^{3/2}f(\nu,T_{\rm
e})^2\ {(T_{\rm \scriptscriptstyle CMBR})}^2 {\sigma_{\rm T}}^2\ (1+z_{\rm
c})^4}\frac{1}{\eta(z)^2} \right] \nonumber \\
& = & D_A^{\: data} \; \eta^{-2},
\end{eqnarray}
where $z_c$ is the galaxy cluster redshift. Therefore, as previously
stressed by Uzan et al.\ (2004), galaxy cluster observations do not
provide the angular diameter distance directly. In principle,
instead of the real angular diameter distance, the measured quantity
is $D_A^{\: data}(z)=D_A(z) \; \eta^2(z)$.

To proceed with our analysis, the quantity $\eta(z)$ defining the
deformed DD relation (see Eq. (\ref{receta})) is parametrically
described by  two different expressions (Holanda, Lima \& Ribeiro
2010; 2011)

\begin{equation}\label{param}
i) \,\,\,\eta (z) = 1 + \eta_{0} z, \, \, \, \, \, \,ii)\,\,\, \eta
(z) = 1 + \eta_{0}z/(1+z).
\end{equation}
\noindent   The first expression is a continuous and smooth
one-parameter linear expansion, whereas the second one includes a
possible epoch-dependent correction that avoids the divergence at
extremely high z.  Naturally, one may argue that these relations were
not derived from first principles. However, we stress that these 
expressions are very simple and have several advantages such as
a manageable one-dimensional phase space and a good
sensitivity to observational data. Clearly, the second
parametrization can also be rewritten as $\eta (z) = 1 +
\eta_{0}(1-a)$, where $a(z)= (1 + z)^{-1}$ is the cosmic-scale
factor. It represents an improvement with respect to the linear
parametrization, since the DD relation becomes bounded regardless of
the redshift values. Potentially, it will become more useful once
higher redshift cluster data are made available. 

The above parametrizations are clearly inspired by similar
expressions for the $\omega(z)$-equation of state parameter in
time-varying dark energy models (see, for instance,  Padmanabhan \&
Choudury 2003; Linder 2003; Cunha, Marassi \& Santos 2007; Silva,
Alcaniz \& Lima 2007). In the limit of very low redshifts ($z<<1$),
we have $\eta = 1$ and $D_{L} = D_{A}$ as should be expected, and,
more important for our subsequent analysis, the value $\eta_0=0$
must be favored by the Etherington principle. In other words, for a
given data set, the likelihood of $\eta_0$ must be peaked at
$\eta_0=0$, in order to satisfy the Etherington theorem. It should be
remarked that Gon\c{c}alves, Holanda \& Alcaniz (2011) also adopted
these expressions to explored the DD relation by using  
observations of gas mass fractions of galaxy clusters, whereas
Lima, Cunha \&  Zanchin (2011) showed how to derive them from
first principles based on  possible theoretical modifications of the
luminosity distance (without refraction effects).

\section{Galaxy cluster samples }
\label{sec:sample}

Below, the physical constraints encoded in the possibility
of a deformed DD relation is explored by considering three samples
of angular diameter distances from galaxy clusters obtained by
combining their SZE and X-ray surface brightness observations. The
first is defined by 38 angular diameter distances from galaxy
clusters in the redshift range $0.14\leq z \leq 0.89$, as given in
the Bonamente et al.\ (2006) sample, where the hydrostatic equilibrium
model and spherical symmetry was considered to describe the galaxy
clusters. In order to construct a realistic model for the cluster
gas distribution and include the possible presence of the
cooling flow, these authors modeled the gas density with a function
whose form is given below (Bonamente et al.\ 2006; La Roque et al.\
2006),
\begin{eqnarray}
\label{density}
n_e(r) = n_{e0} \; \; \; \cdot & & \left[f\left( 1+\frac{r^2}{r_{c1}^2}
\right)^{-\frac{3\beta}{2}}+ \right. \nonumber \\
& & + \; (1-f) \left. \left( 1+\frac{r^2}{r_{c2}^2}
\right)^{-\frac{3\beta}{2}} \right].
\end{eqnarray}
This double $\beta$-model for the density generalizes the single
$\beta$-model profile, introduced by Cavaliere and Fusco-Fermiano
(1976) and the double $\beta$ model proposed by Mohr et al.\ (1999).
It has the freedom of following both the central spike in density
and the more gentle outer distribution. The quantity $n_{e0}$ is the
central density, $f$ governs the fractional contributions of the
narrow and broad components ($0\leq f \leq 1$), $r_{c1}$ and
$r_{c2}$ are the two core radii that describe the shape of the inner
and outer portions of the density distribution and $\beta$
determines the slope at large radii (the same $\beta$ is used for
both the central and outer distributions in order to reduce the
total number of degrees of freedom).

On the other hand, the hydrostatic equilibrium and spherical
symmetry hypotheses result in the condition
\begin{equation}
\frac{dP}{dr}=-\rho_g \frac{d\phi}{dr},
\end{equation}
where $P$ is the gas pressure, $\rho_g$ is the gas density and
$\phi=-G M(r)/r$ is the gravitational potential due to dark
matter and the plasma.  Using the ideal gas equation of state for
the diffuse intracluster plasma, $P=\rho_g k_B T / \mu \; m_p$,
where $\mu$ is the mean molecular weight and $m_p$ is the proton
mass, one obtains a relationship between the cluster temperature and
the cluster mass distribution,
\begin{eqnarray}
\label{eq-hse}
\frac{dT}{dr} &=&-\left( \frac{\mu m_p}{k_B} \frac{d\phi}{dr} +
\frac{T}{\rho_g} \frac{d\rho_g}{dr} \right), \nonumber \\
&=& -\left( \frac{\mu m_p}{k_B}
\frac{G M}{r^2} + \frac{T}{\rho_g} \frac{d\rho_g}{dr} \right).
\end{eqnarray}
{Bonamente et al.\ (2006) combined hydrostatic equilibrium equations
with a dark matter density distribution from Navarro, Frenk and
White (1997),
\begin{equation}
\rho_{\scriptscriptstyle DM}(r)=\mathcal{N} \left[ \frac{1}{(r/r_s) (1+r/r_s)^2}
\right],
\label{NFW}
\end{equation}
where $\mathcal{N}$ is a density normalization constant and $r_s$ is
a scale radius. The parameters of these equations ($n_{e0}$, $f$,
$r_{c1}$, $r_{c2}$, $\beta$, $\mathcal{N}$ and $r_s$) were obtained
by calculating the joint likelihood $\mathcal{L}$ of the X-ray and
SZE data in a Markov chain Monte Carlo method (Bonamente et al.\
2004). Summarizing, the cluster plasma and dark matter distributions
were analyzed assuming hydrostatic equilibrium model and spherical
symmetry, thereby accounting for radial variations in density,
temperature and abundance. }

The second sample is formed by 25 galaxy clusters in the redshift
range $0.023 \leq z \leq 0.8$ compiled by De Filippis et al.\
(2005). These authors re-analyzed archival X-ray data of the
XMM-Newton and Chandra satellites of two samples for which combined
X-ray and SZE analysis have already been reported using an
isothermal spherical $\beta$-model. One of the samples, compiled
previously by Reese et al.\ (2002), is a selection of 18 galaxy
clusters distributed over the redshift interval $0.14 < z < 0.8$.
The other one, the sample of Mason et al.\ (2001), has seven clusters
from the X-ray limited flux sample of Ebeling et al.\ (1996). In
this way, De Filippis et al.\ (2005) used an isothermal elliptical
$\beta$-model and an isothermal spherical $\beta$ model to obtain
$D_A(z)$ measurements for these galaxy clusters samples. As
discussed by De Filippis et al.\ (2005), the choice of circular rather
than elliptical $\beta$ model does not affect the resulting 
central surface brightness or Sunyaev-Zeldovich decrement, the slope
$\beta$ differs slightly between these models, but
significantly different values for core radius are obtained. The
result was that the core radius of the elliptical $\beta$-model is
bigger than that of the spherical $\beta$ model (see Fig. 1 in their
paper). In a first approximation it was found that $\theta_{ell} =
\frac{2e_{proj}}{1+e_{proj}}\theta_{circ}$, where $e_{proj}$ is the
axial ratio of the major to the minor axes of the projected
isophotes. Since $D_A \propto \theta^{-1}_c$, angular diameter
distances obtained by using an isothermal spherical $\beta$-model
are overestimated compared with those from the elliptical $\beta$-model.

For the isothermal elliptical $\beta$-model De Filippis et al.\
(2005) used a general triaxial morphology  to describe the intra-cluster
medium. They obtained
\begin{equation}
\label{eqsz2} \Delta T_{\rm SZ} = \Delta T_0 \left[ 1+
\frac{{\theta_{1}}^2+ {(e_{\rm proj})}^2
{\theta_{2}}^2}{{(\theta_{c,\rm proj})}^2} \right]^{1/2-3\beta/2},
\end{equation}
where
\begin{eqnarray}
\label{eq:sze3} \Delta T_0 &\equiv &T_{\rm \scriptscriptstyle CMBR}
 \: f(\nu, T_{\rm
e}) \frac{ \sigma_{\rm T} \: k_{\rm B} \: T_{\rm e}}{m_{\rm e} \:
c^2}n_{e0}
\sqrt{\pi} \times \nonumber \\
& & \times \frac{D_{\rm A} \: \theta_{\rm c,proj}}{h^{3/4}}
\sqrt{\frac{e_1 e_2}{e_{\rm proj}}} \; g\left(\beta/2\right),
\end{eqnarray}
$g(\alpha)$ is given by equation (\ref{galfa}),
$D_{\rm A}$ is the angular diameter distance to the cluster,
$\theta_i \equiv x_{i,\rm obs}/D_{\rm A}$ is the projected angular
position (on the plane of the sky) of the intrinsic orthogonal
coordinate $x_{i,\rm obs}$, $h$ is a function of the cluster shape
and orientation, $e_{\rm proj}$ is the axial ratio of the major to
the minor axes of the observed projected isophotes and
$\theta_{c,\rm proj}$ is the projection on the plane of the sky
(p.o.s.) of the intrinsic angular core radius.

Similarly, the X-ray surface brightness $S_{X0}$ can be written as
follows,
\begin{equation}
S_X = S_{X0} \left[ 1+ \frac{{\theta_{1}}^2+{(e_{\rm proj})}^2
{\theta_{2}}^2}{{(\theta_{c,\rm proj})}^2} \right]^{1/2-3 \beta},
\label{eq:sxb1}
\end{equation}
where the central surface brightness $S_{X0}$ reads

\begin{equation}
\label{eq:sxb2} S_{X0} \equiv \frac{ \Lambda_{eH} D_A^{2}\
(\mu_e/\mu_H)}{4 \sqrt{\pi} \: D_L^{2}} \; n_{e0}^2 \;  \frac{D_{\rm
A} \: \theta_{\rm c,proj}}{h^{3/4}}\sqrt{\frac{e_1 e_2}{e_{\rm
proj}}}\ \; g(\beta),
\end{equation}
and $\mu$ is the molecular weight, given by $\mu_i\equiv
\rho/n_im_p$.

By eliminating $n_{\rm e0}$ in the equations above and assuming the
DD relation as valid, i.e., $D_A^{\: data}(z)=D_A(z)$, De Filippis
et al.\ (2005) obtained the observational quantity as written below
\begin{eqnarray}
\label{eq:obl7}
D_{A}(z) &= & \left[ \frac{{\Delta T_0}^2}{S_{\rm X0}}
\left( \frac{m_{\rm e} \: c^2}{k_{\rm B} \: T_{e0} } \right)^2
\frac{g\left(\beta\right)}{{g(\beta/2)}^2\ \theta_{\rm c,proj}}
\right] \times 
\nonumber \\
& \times &  \left[ \frac{\Lambda_{eH0}\ ( \mu_e/\mu_H)}{4
\pi^{3/2}f(\nu,T_{\rm e})^2\ { \left( T_{\rm \scriptscriptstyle CMBR}
\right) }^2\ {\sigma_{\rm T}}^2\ (1+z_{\rm c})^4} \right].
\end{eqnarray}
The slope $\beta$ of the profile and the projected core radius
$\theta_{c,proj}$ were obtained by fitting the cluster surface
brightness with an elliptical 2-D $\beta$ model.  {For the
isothermal spherical $\beta$ model description, De Filippis et al.\
(2005) considered the usual Eq. (8) and obtained angular distances
with Eq. (14) and $\eta =1$.}

In Fig.\ 1 we plot the galaxy cluster samples. In Fig.\ 1a the
filled circles (blue) and open  squares (black) with the associated
error bars (only statistical errors) stand for the De Filippis et
al.\ (2005) isothermal elliptical $\beta$ model and  isothermal
spherical $\beta$ model, respectively. In Fig.\ 1b we show the
sample of Bonamente et al.\ (2006), where a non-isothermal spherical
double $\beta$ model was used to describe the galaxy clusters (red
filled squares).

\section{Obtaining the shape of galaxy clusters by using the DD relation
as constraint }\label{sec:meth1}

Many studies about the intra-cluster gas and dark matter (DM)
distribution in galaxy clusters have been limited to the standard
spherical geometry (Reiprich \& Boringer 2002; Bonamente et al.\
2006; Shang, Haiman \& Verdi 2009). However, in the past few years
observations of galaxy clusters based on {\it Chandra } and {\it
XMM} data have shown that generally clusters exhibit elliptical surface
brightness maps. Simulations have also predicted that DM halos show
axis ratios typically on the order of $\approx 0.8$ (Wang \& White
2009), thereby disproving the spherical geometry assumption. In this
line, the first determination of the intrinsic three-dimensional
(3D) shapes of galaxy clusters was presented by Morandi, Pedersen \&
Limousin (2010) by combining X-ray, weak-lensing and strong-lensing
observations. Their methodology was applied to the galaxy cluster
MACS J1423.8+2404 and they found a triaxial galaxy
cluster geometry with DM halo axial ratios $1.53 \pm 0.15$ and $1.44
\pm 0.07$ on the plane of the sky and along the line of sight,
respectively.

Bearing in mind these results, we propose a new method to access the
galaxy cluster morphology by taking the validity of the DD relation
as a constraint. The idea is very simple. Beacuse the samples shown in
\S\ref{sec:sample} were compiled under different geometric assumptions,
we confront these underlying hypotheses with the validity of the DD
relation. In principle, this kind of result provides an interesting
example of how a cosmological (global) condition correlates to the
local physics. { In the application,  one should also keep in mind
that a deformed  DD relation as given by Eq. 3 naturally induces a
more general result for the angular diameter distance from galaxy
clusters via SZE and X-ray technique (see Eq. 14)}
\begin{equation}
D^{\: data}_{A}(z)=D_{A}(z) \: \eta^{2}.
\end{equation}
In this line, our aim is to estimate the $\eta_{0}$ parameter for
each galaxy cluster sample for both parameterizations of  $\eta(z)$
as given by Eq. (\ref{param}). In our analyses (Holanda, Lima \&
Ribeiro 2011), $D_A(z)$ for each galaxy clusters} is obtained from the
WMAP (seven years) results by fixing the conventional flat $\Lambda$CDM
model. The parameters of the simplest six-parameter $\Lambda$CDM model
were recently determined by  Komatsu  et al.\ (2011) by using the
combination of seven-year data from WMAP with the latest distance
measurements from baryon acoustic oscillations (BAO) in the
distribution of galaxies (Percival et al.\ 2010) and the Hubble
parameter $H_{0}$ measurement presented by Riess et al.\ (2009). The
basic results are $\Omega_{\Lambda}=0.725\pm0.016$ and
$h=0.702\pm0.014$, and no  convincing evidence for deviations from
the minimal cosmic concordance model has been established.

The theoretical angular diameter distance can be written as (Lima et
al.\ 2003; Cunha et al.\ 2007)
\begin{equation}
{{D}}_{A}(z;h,\Omega_m) = \frac{3000h^{-1}}{(1 +
z)}\int_{o}^{z}\frac{dz'}{{\cal{H}}(z';\Omega_m)}\,\mbox{Mpc},
\label{eq1}
\end{equation}
where $h=H_0/100$ km s$^{-1}$ Mpc$^{-1}$ and the dimensionless
function ${\cal{H}}(z';\Omega_m)$ is given by
\begin{equation}
{\cal{H}} = \left[\Omega_m(1 + z')^{3} + (1 -\Omega_m)\right]^{1/2}.
\label{eq2}
\end{equation}

Now, in order to constrain $\eta_0$, let us evaluate the likelihood
distribution function,
$e^{-\chi^{2}/2}$, where
\begin{equation}
\label{chi2} \chi^{2} = \sum_{z}\frac{{\left\{ \; {[\eta(z)]}^2 -
{[\eta_{obs}(z)]}^2 \; \right\} }^{2}}{\sigma^2_{{\eta_{obs}}} },
\end{equation}
with $[\eta^2_{obs}(z)] = D^{\: data}_{A}(z)/D_{A}(z)$ and
\begin{equation}
\sigma^2_{\eta_{obs}}=\left[\frac{1}{D_A(z)}\right]^2\sigma^{2}_{data} + 
\left[\frac{D_A^{data}}{D^2_A(z)}\right]^2 \sigma^{2}_{\scriptscriptstyle WMAP},
\end{equation} 
where $\sigma^2_{\scriptscriptstyle WMAP}$  
is the  error in $D_A(z)$ associated to cosmological parameters. The common
statistical error contributions  for galaxy clusters are SZE point
sources $\pm 8$\%, X-ray background $\pm 2$\%, Galactic N$_{H}$
$\leq \pm 1\%$, $\pm 15$\% for cluster asphericity, $\pm 8$\%
kinetic SZ and for CMBR anisotropy $\leq \pm 2\%$. On the other
hand, the estimates for systematic effects are SZ
calibration $\pm 8$\%, X-ray flux calibration $\pm 5$\%, radio halos
$+3$\%, and X-ray temperature calibration $\pm 7.5$\%.  Indeed, one
may show that typical statistical errors amount to nearly
$20$\%, in agreement with other works (Mason et al.\ 2001; Reese et
al.\ 2002, 2004),  while for systematics we also find typical errors
around + 12.4\% and - 12\%  (see also Table 3 in Bonamente et al.\
2006). In the present analysis we have combined the statistical and
systematic errors in quadrature for the galaxy clusters
($\sigma^{2}_{data}=\sigma^{2}_{stat}+\sigma^{2}_{syst}$).
We note that in our $\chi^2$ statistical analysis the asymmetric
uncertainties present in the Bonamente et al.\ (2006) and De Filippis
et al.\ (2005) samples  were symmetrized by using the D'Agostini
(2004) method.

In Figs.\ 2a and 2b we plot the likelihood distribution function for
the De Filippis et al.\ (2005) and Bonamente et al.\ (2006) samples.
The $\eta_0$ values and their errors (statistical +
systematic errors) of our analysis are given below (see Table
\ref{table1}).

\begin{table*}[ht!]
\caption{{The $\eta_0$ values and their errors (statistical +
systematic errors) of our first analysis (section \ref{sec:meth1}).
}}
\begin{center}
\begin{tabular} {|c|c|c|}
\hline\hline  Isothermal elliptical $\beta$ model  x $\Lambda$CDM
(WMAP7)& $\chi^{2}/d.o.f$
\\ \hline \hline
$\eta_{0}$ (Linear case) = $ -0.056\pm - 0.10$ (1$\sigma$)& 23.52/24    \\
$\eta_{0}$ (Non-linear case) = $-0.088 \pm 0.14$ (1$\sigma$)&22.56/24 \\
\hline\hline  Isothermal spherical $\beta$ model  x $\Lambda$CDM (WMAP7)& $\chi^{2}/d.o.f$
\\ \hline \hline
$\eta_{0}$ (Linear case) = $0.19 \pm 0.12$ (1$\sigma$) & 20.16/24    \\
$\eta_{0}$ (Non-linear case) = $0.28 \pm 0.18$ (1$\sigma$)& 19.44/24  \\
\hline\hline Non-isothermal spherical $\beta$ model  x $\Lambda$CDM (WMAP7)&$\chi^{2}/d.o.f$
\\ \hline \hline
$\eta_{0}$ (Linear case) =  $-0.11 \pm 0.06$ { (1$\sigma$)} & 29.6/37\\
$\eta_{0}$ (Non-linear case) = $-0.16 \pm 0.08$ { (1$\sigma$)} &29.23/37 \\
\hline \hline
\end{tabular} \label{table1}
\end{center}
\end{table*}

For an isothermal elliptical $\beta$ model, we see that
the angular diameter distances from the De Filippis et al.\ (2005)
sample provide an excellent fit and agree with the DD relation
at 1$\sigma$ confidence level. On the other hand, although it agrees
with the DD relation at 2$\sigma$ c.l., the analysis based on the
isothermal spherical $\beta$ model leads to $\eta_0$ values higher
than those from the elliptical $\beta$ case. Since in this case $\eta_0
> 0$, the departures of the DD relation validity indicates that the
estimated angular distances with the spherical $\beta$ model are
overestimated with respect to those from the conventional
flat $\Lambda$CDM (WMAP7). For the Bonamente et al.\ (2006)
sample (non-isothermal spherical double $\beta$ model) we also see
that the strict validity of the reciprocity relation is only
marginally compatible. The relative situation is not modified even
when only clusters with $z>0.1$ are considered in the De Filippis et
al.\ (2005) sample. In such a situation, we obtain $\eta_{0} =
-0.044^{+ 0.1}_{- 0.1}$ ($\chi^2/d.o.f.= 15.9/17$) for the linear
parametrization, and $\eta_{0} = -0.07^{+ 0.14}_{- 0.14}$
($\chi^2/d.o.f.= 15.6/17$) within 1$\sigma$ in the non-linear case
for elliptical description and $\eta_{0} = 0.186^{+ 0.11}_{- 0.1}$
($\chi^2/d.o.f.= 10.9/17$) for the linear parametrization, and
$\eta_{0} = 0.274^{+ 0.145}_{- 0.145}$ ($\chi^2/d.o.f.= 10.7/17$)
within 1$\sigma$ in the non-linear case for spherical description.

Therefore, we found no evidence for a distance-duality
violation for the elliptical De Filippis et al.\ sample (2005).
However, the  same kind of analysis contradicts 
the spherical symmetry hypothesis assumed in the Bonamente et
al.\ (2006) sample and in the De Filippis et al.\ (2005)
sample when a spherical geometry is assumed. These results are very
interesting since they show how important the choice of geometry is
in describing the clusters to obtain their distances
through SZE + X-ray measurements. We also see that the
non-isothermal assumption of Bonamente et al.\ (2006) in their
spherical description was not sufficient to satisfy the validity
of the DD relation in the $\Lambda$CDM framework (WMAP7, Komatsu
et al.\ 2011).

We recall that Bonamente et al.\ (2006) used three different models
to describe the same 38 galaxy clusters: (i) the non-isothermal
spherical double $\beta$  model already discussed, (ii) an isothermal
spherical $\beta$ model, and (iii) the isothermal spherical $\beta$
model excluding the central 100 kpc from the X-ray data. For the sake
of completeness, we also obtained the $\eta_0$ values for the last
two descriptions. For an isothermal spherical $\beta$ model with
linear and non-linear parametrizations we found  $\eta_{0} = -0.09^{+
0.05}_{- 0.06}$ ($\chi^2/d.o.f. = 48.48/37$) and $\eta_{0} =
-0.13^{+ 0.08}_{- 0.09}$ ($\chi^2/d.o.f. = 47.56/37$), respectively. 
In the isothermal spherical $\beta$ model excluding the central 100
kpc from the X-ray data we obtained $\eta_{0} = -0.14^{+ 0.07}_{- 0.07}$
($\chi^2 /d.o.f. = 47.81/37$) for the linear parametrization, and
$\eta_{0} = -0.2^{+ 0.09}_{- 0.11}$ ($\chi^2 /d.o.f. = 48.40/37$) in
the non-linear case. All these determinations are at a 1$\sigma$
confidence level. 

Furthermore, Bonamente et al.\ (2006) also determined the Hubble
constant $H_0$ for each model. By assuming a flat $\Lambda$CDM model
($\Omega_M =0.3$ and $\Omega_{\Lambda}$=0.7), the corresponding values
at 1$\sigma$ c.l. are $H_0=76.9^{+3.9}_{-3.4}$, $H_0=73.7^{+4.6}_{-3.8}$
and $H_0=77.6^{+4.8}_{-4.3}$. More recently, Cunha, Marassi \& Lima
(2007) obtained $H_0=74\pm3.5$ through a joint analysis with BAO by
using the De Filippis et al.\ sample (2005), thereby alleviating the
tension between between SZ + X-ray technique and the CMB + BAO
determination of $H_0$. However, since the Hubble constants obtained
from  all these models agree at 1$\sigma$, one may conclude that the
$H_0$ value is not useful to decide which galaxy cluster model is
more realistic.

\section{Testing the DD relation with galaxy clusters and SNe Ia }
\label{sec:meth2}

In this section  we apply a model-independent cosmological test for
the DD relation based on the three samples defined in  section 3. For
$D_L$ we considered two subsamples of SNe type Ia taken from Tables
I and II of the Hicken et al.\ (2009) (Constitution data ), whereas
values for $D_A$ are provided by the three samples of galaxy
clusters discussed above. The SNe Ia redshifts of each subsample
were carefully chosen to coincide with those of the associated
galaxy cluster sample ($\Delta z<0.005$), thereby allowing a direct
test of the DD relation. For a given pair of data sets (SNe Ia,
galaxy clusters), one should expect a likelihood of $\eta_0$ peaked
at $\eta_0=0$, in order to satisfy the DD relation. Moreover, in
our approach the data do not need to be binned  as
assumed in some analyses involving the DD relation (see, for
instance, De Bernardis, Giusarma \& Melchiorri 2006).

In Fig.\ 3a we plot $D_A$ multiplied by $(1+z_{cluster})^{2}$ from
the galaxy clusters sample compiled from the De Filippis et al.\
sample (2005) and $D_{L}$ from our first SNe Ia subsample. In Fig.\ 3b we plot
the subtraction of redshift between clusters and SNe Ia. We see that
the biggest difference is $\Delta z \approx 0.01$ for three clusters
(open squares), while for the remaining 22 clusters we have $\Delta
z < 0.005$. In order to avoid the corresponding bias, we removed the
three clusters from all subsequent analyses so that $\Delta
z < 0.005$ for all pairs.

Similarly, in Fig.\ 4a we plot $D_A$ multiplied by
$(1+z_{cluster})^{2}$, but now for the Bonamente et al.\  sample
(2006) and $D_{L}$ from our second SNe Ia subsample. In Fig.\ 4b we
display the redshift subtraction between clusters and SNe Ia. Again,
we see that for 35 clusters $\Delta z < 0.005$. The biggest
difference is again $\Delta z \approx 0.01$ for three clusters, and, for
consistency, these were also removed from our analysis.

Let us now estimate the parameter $\eta_{0}$ for each sample and
the two parametrizations defined by Eq.\ (\ref{receta}). We
recall that in general the SZE + X-ray surface brightness
observations technique does not produce $D_{A}(z)$, but $D^{\:
data}_{A}(z)=D_{A}(z) \; \eta^{2}$. Consequently, if one wishes to
test Eq.\ (1) with SZE + X-ray observations from galaxy clusters,
the angular diameter distance must be replaced by $D^{\: data}_{A}(z) \:
\eta^{-2}$ in Eq. (3). In this way, we obtained $\eta(z)=D^{\:
data}_{A}(z)(1+z_{cluster})^{2}/D_{L}(z)$.

Following standard procedure, the likelihood estimator is determined
by a $\chi^{2}$ statistics
\begin{equation}
\label{chi22} \chi^{2} = \sum_{z}\frac{{\left[\eta(z) -
\eta_{obs}(z) \right] }^{2}}{\sigma^2_{\eta_{obs}} },
\end{equation}
where $\eta_{obs}(z) = (1+z_{cluster})^{2}D^{\:
data}_{A}(z)/D_{L}(z)$ and
\begin{eqnarray}
\sigma^2_{\eta_{obs}} &=& \left[\frac{(1+z_{cluster})^{2}D^{data}_A}
{D^2_L}\right]^2 \sigma^2_{D_L}+ \nonumber \\
& &+ \left[\frac{(1+z_{cluster})^{2}}
{D_L}\right]^2\sigma^2_{D^{data}_A}.
\end{eqnarray}
As in the previous section, we combined the statistical and systematic
errors in quadrature for the angular diameter distance from galaxy
clusters, and, as we remarked above, the asymmetric error bars were
treated by the D'Agostini (2004) method.

On the other hand, after nearly 500 discovered SNe Ia, the
constraints on the cosmic parameters from luminosity distance are
now limited by systematics rather than by statistical errors. In
principle, there are two main sources of systematic uncertainties in
SNe Ia observations, which are closely related to photometry and
possible corrections for light-curve shape (Hicken at al.\ 2009).
However, at the moment it is not so clear how one is to estimate the
overall systematic effects for these standard candles (Komatsu et al.\
2011), and, therefore, we exclude them from the following
analysis. The basic reason is that systematic effects from galaxy
clusters seem to be larger than those of SNe observations, but
their inclusion does not affect the results validity of the
distance-duality relation very much. In this way, following
Holanda, Lima \& Ribeiro (2010), we included only statistical
errors of the SNe Ia magnitude measurements.

In Figs.\ 5a and 5b we plot the likelihood distribution function for
each sample. The $\eta_0$ values with their errors (statistical and
systematic errors) are provided in Table \ref{table2}. Clearly, the
confrontation between the angular diameter distances from the elliptical
De Filippis et al.\ (2005) sample with SNe Ia data disagree 
moderately with the reciprocity relation (the DD relation is
marginally satisfied in $2\sigma$) and the $\eta_0$ values are
mostly negative. Since $(1+z_{cluster})^{2}D^{\:
data}_{A}(z)/D_{L}(z) \propto 1+ \eta_{0}z$, negative $\eta_0$
values indicate luminosity distances overestimated with respect to
the angular diameter ones. This tension between Sne Ia and the
elliptical De Filippis et al.\ (2005) sample arises because SNe Ia
samples prefer universes with a higher $\Omega_{\Lambda}$ value than
those of the WMAP7 results. Indirectly, our analysis support the tension
found by Wei (2010), who used SNe Ia constitution data and observations
of the CMBR anisotropy plus BAO separately to constrain
the $\omega_a$ parameter of the dark energy equation of state given
by $\omega=\omega_0 + \omega_a z/(1+z)$. The SNe Ia constitution
data indicated a strong negative $\omega_a$ ($\omega_a \approx -11$)
and the phantom energy divide line ($\omega <-1$) could be crossed
at recent redshifts. Furthermore, the constitution data disagreed 
not only with the CMB and BAO observations, but also with other SNe
Ia datasets such as Davis (Davis et al.\ 2007) and SNLS (Astier et
al.\ 2006). On the other hand, when an isothermal spherical $\beta$
model is used, the DD relation is satisfied in $1\sigma$. As
discussed earlier, this concordance occurs because an isothermal
spherical $\beta$ model yields angular distances overestimated 
compared to the elliptical model and WMAP7 results. For the
Bonamente et al.\ (2006) sample, where a non-isotermal spherical
double $\beta$ model was assumed to describe the clusters, we see
that the DD relation is not obeyed even at 3$\sigma$. We note that
our results do not change if one replaces $z_{cluster}$
by $z_{SNe}$, where  $z_{SNe}$ is the redshift of SNe Ia. This 
shows that a $\Delta z < 0.005$ is sufficient to implement our
direct test.

\begin{table*}[ht!]
\caption{{The $(\eta_0)$ values and their errors (statistical +
systematic errors) of our second analysis (section \ref{sec:meth2}).
}}
\begin{center}
\begin{tabular} {|c|c|c|}
\hline\hline  Isothermal elliptical $\beta$ model  x SNe Ia&
$\chi^{2}/d.o.f$
\\ \hline \hline
$\eta_{0}$ (Linear case) = $ -0.28 \pm 0.21$ { (1$\sigma$)}& 21.2/21    \\
$\eta_{0}$ (Non-linear case) = $-0.43 \pm 0.29$ { (1$\sigma$)}&21/21 \\
\hline\hline  Isothermal spherical $\beta$ model  x SNe Ia & $\chi^{2}/d.o.f$
\\ \hline \hline
$\eta_{0}$ (Linear case) = $0.14 \pm 0.26$ { (1$\sigma$)}& 17.01/21    \\
$\eta_{0}$ (Non-linear case) = $0.20 \pm 0.36$ { (1$\sigma$)} &16.80/21  \\
\hline\hline Non-isothermal spherical $\beta$ model  x SNe Ia&$\chi^{2}/d.o.f$
\\ \hline \hline
$\eta_{0}$ (Linear case) = $-0.39 \pm 0.11${ (1$\sigma$)}  & 30.26/34\\
$\eta_{0}$ (Non-linear case) = $-0.61 \pm 0.16$ { (1$\sigma$)}& 28.9/34 \\
\hline \hline
\end{tabular} \label{table2}
\end{center}
\end{table*}
In this context, we recall that Li, Wu and Yu (2011) rediscussed
this independent cosmological test using the
latest Union2 SNe Ia data and the angular diameter distances from
galaxy clusters. However, they removed six and twelve data points,
respectively, from the De Filippis et al.\ (2005) and Bonamente et
al.\ (2006) samples and obtained a more serious violation of the DD
relation. These authors also reexamined the DD relation by
postulating two more general parameterization forms, namely $\eta(z)
= \eta_0 + \eta_1 z$ and $\eta(z) = \eta_0 + \eta_1 z/(1+z)$, and
they found that consistencies between the observations and the DD
relation are markedly improved for both samples of galaxy clusters.
Nair, Jhingan and Jain (2011) also discussed the strict validity of
the DD relation using the latest Union2 SNe Ia data and the angular
diameter distances from galaxy clusters, FRIIb radio galaxies and
mock data. They proposed six different (one and two indices)
parametrizations (including, as particular cases, those adopted
by Holanda et al.\ 2010) in an attempt to determine a possible
redshift variation of the DD relation. As physically expected, their
results depend both on the specific parametrization and the data
samples considered. In particular, they conclude that one index
parameterization, namely $\eta_{V}=\eta_{8}/(1 + z)$ and
$\eta_V=\eta_{9} \exp \{ [z/(1 + z)]/(1 +z) \}$, do not support
the DD relation  for the given data set.

More recently, Meng, Zhang and Zhan (2011) also reinvestigated the
model-independent test by comparing two different methods and several
parametrizations (one and two indices) for $\eta(z)$. Their basic
conclusion agrees with our studies in the sense that the
triaxial ellipsoidal model is suggested by the model-independent
test at 1$\sigma$ while the spherical $\beta$ model can only be
accommodated at a $3\sigma$ confidence level. As we shall see,
this uncertainty can also be robustly resolved by considering a
joint analysis involving the two treatments discussed here (see
section 7). In principle, since all results somewhat depend on
the assumed $\eta(z)$ form, it is important to understand the
effect of different parametrizations. However, we recall that at
extremely low redshifts the DD relation reduces to unity since
$D_L(z)\equiv D_A(z)$, and, therefore, the constant parameter
appearing in the proposed $\eta(z)$ expressions should be fixed to
unity.

At this point it is interesting to know  the influence of SNe Ia
light-curve fitters on the model-independent test of the distance
duality relation and some of its consequences for the local physics.
This topic has attracted a lot of attention in the recent literature
(Kessler et al.\ 2009; Bengochea 2010, Hao Wei 2011),  and its
connection with the DD relation and galaxy cluster geometry it will
be discussed next.

\section{SNe Ia light-curve fitters, DD relation and cluster geometries }
\label{sec:meth3}

In the previous section, we tested the DD relation by using
luminosity distances from SNe Ia based on the Constitution sample
and the angular diameter distances from three galaxy clusters
samples obtained by their SZE and X-rays surface brightness
measurements. However, in the Constitution compilation different SNe
Ia samples were analyzed by four light-curve fitters (SALT, SALT2,
MLCS31 and MLCS17) to test consistency and systematic differences:
397 SNe Ia with SALT, 351 SNe Ia with SALT2, 366 SNe Ia with MLCS31
and 372 SNe Ia with MLCS17. The four fitters were seen to be
relatively consistent with the light-curve-shape and color
parameters, but that still leaves room for improvement, and they
provide a considerable amount of systematic uncertainty to any
analysis. In this way, it would be interesting to investigate the
influence of different SNe Ia light-curve fitters on our previous
results. Unfortunately, it was not possible to find the same SNe's
Ia used here in section \ref{sec:meth2} in the Constitution
sample, but analyzed with the four methods for a direct comparison.

On the other hand, Kessler et al.\ (2009) presented the Hubble
diagram for 103 SNe Ia with redshifts $0.04 < z < 0.42 $, discovered
during the first season of the Sloan Digital Sky Survey-II
(SDSS-II). These data filled the redshift desert between low- and
high-redshift SNe Ia of the previous SNe Ia surveys. In addition, these
authors included a comprehensive and consistent reanalysis of other
datasets (ESSENCE, SNLS, HST), resulting in a combined sample of 288
SNe Ia. Reanalysis was performed by using two light-curve fitters:
SALT2 (Guy et al.\ 2007) and MLCS2K2 (Jha et al.\ 2007). MLCS2K2
calibration uses a nearby training set of SNe Ia assuming a close to
linear Hubble law, while SALT2 uses the whole data set to calibrate
empirical light curve parameters, and a cosmological model must be
assumed in this method. Typically a $\Lambda$CDM or a $\omega$CDM
model is assumed. Consequently, the SNe Ia distance moduli obtained
with SALT2 fitter retain a degree of model dependence (Bengochea
2011). Kessler et al.\ (2009) combined these 288 SNe Ia  with
measurements of baryon acoustic oscillations from the SDSS luminous
red galaxy sample and with cosmic microwave background temperature
anisotropy measurements from the WMAP for estimating the
cosmological parameters $\omega$ and $\Omega_{M}$ by assuming a
spatially flat cosmological model. The results
from these two light-curve fitters disagreed: $\omega = -0.76 \pm
0.07$(stat)$\pm 0.11$(syst), $\Omega_{M}=0.307 \pm 0.019$(stat)$\pm
0.023$(syst) using MLCS2K2 and $\omega = -0.96 \pm 0.06$(stat)$\pm
0.12$(syst), $\Omega_{M}=0.265 \pm 0.016$(stat)$\pm 0.025$(syst)
using SALT2. This discrepancy raised the question of which method
is more reliable because it is not possible to definitively determine
from the current data that either method is better or incorrect. The
overall conclusion was that the cosmological parameter $\omega$ lies
between $-1.1$ and $-0.7$ (Kessler et al.\ 2009). Exploring
nonstandard cosmology scenarios, Sollerman et al.\ (2009) found that
more exotic models provided a better fit to the SNe Ia data than the
$\Lambda$CDM model when the MLCS2K2 light-curve fitter was used.
However, when the SN Ia data were analyzed by using SALT2 light-curve
fitter, the standard cosmological constant model agreed better
to those data.

In this section, by using the SNe Ia data from SDSS-II (Kessler et
al.\ 2009), we explore the influence of the two different SNe Ia
light-curve fitters on our model-independent cosmological test for
the DD relation.  For $D_L$ we considered two subsamples of SNe Ia
taken from SDSS-II (2007) where, again, the SNe Ia redshifts of each
subsample were carefully chosen to coincide with those of the
associated galaxy cluster sample ($\Delta z<0.005$), allowing a direct
test of the DD relation. Furthermore, in this case, each subsample of
SNe Ia was analyzed by the MLCS2K2 and SALT2 light-curve fitters
separately. For $D_A$ we used the angular diameter distance samples
of galaxy clusters discussed above: the isothermal elliptical and
spherical $\beta$ model samples from De Filippis et al.\ (2005) and
the non-isothermal spherical double $\beta$ model sample from
Bonamente et al.\ (2006).

In Fig.\ 6a we plot the subtraction of redshift between clusters and
SNe Ia for the De Filippis et al.\ (2005) sample. We see that the
biggest difference is $\Delta z \approx 0.007$ for three clusters (open
squares), whereas for the remaining 22 clusters we have $\Delta z <
0.005$. In order to avoid the corresponding bias, the three clusters
were removed from all analyses presented here so that $\Delta z <
0.005$ for all pairs. In Fig.\ 6b we display the redshift
subtraction between clusters and SNe Ia for the Bonamente et al.\
(2006) sample. In this case we see that for 33 clusters $\Delta z <
0.005$. The biggest difference is  $\Delta z \approx 0.015$  for five
clusters, and, for consistency, they were also removed from our
analysis.

In Figs. 7a and 7b we plot the likelihood distribution function for
the elliptical and spherical De Filippis et al.\ (2005) samples,
respectively. The results for Bonamente et al.\ (2006) sample are
shown in Fig.\ 8. Following Kessler et al.\ (2009) and Sollerman et
al.\ (2009), we added in our analyses an additional intrinsic
dispersion of 0.16 mag to the uncertainties output by the MLCS2K2
light-curve fitter and 0.14 mag for the SALT2 light-curve fitter. In
Table \ref{table3} we display our results, where the errors are in
2$\sigma$ (statistical plus systematic errors).

\begin{table*}[ht!]
\caption{{The $\eta_0$ values for our third analysis (section
\ref{sec:meth3}). { All errors are at $1\sigma$}.  }}
\begin{center}
\begin{tabular} {|c|c|c|c|c|}
\hline\hline  Isothermal elliptical $\beta$ model & SALT2&
$\chi^{2}/d.o.f$& MLCS2K2 & $\chi^{2}/d.o.f$
\\ \hline \hline
$\eta_{0}$ (Linear case)& $-0.24 \pm 0.24$& 19.74/21 & $-0.43 \pm 0.21$&  21.2/21 \\
$\eta_{0}$ (Non-linear case)& $-0.34 \pm 0.34$&19.53/21 & $-0.63 \pm 0.32$&20.8/21 \\
\hline\hline  Isothermal spherical $\beta$ model  & SALT2& $\chi^{2}/d.o.f$&
MLCS2K2 & $\chi^{2}/d.o.f$
\\ \hline \hline
$\eta_{0}$ (Linear case) & $0.27 \pm 0.27$ & 16/21 & $0.01 \pm 0.34$&  15.7/21 \\
$\eta_{0}$ (Non-linear case) & $0.4 \pm 0.4$ & 15.7/21 & $0.07 \pm 0.35$& 15.5/21 \\
 \hline\hline Bonamente et al.\ Sample & SALT2&$\chi^{2}/d.o.f$ &  MLCS2K2&$\chi^{2}/d.o.f$
\\ \hline \hline
$\eta_{0}$ (Linear case) & $-0.22 \pm 0.14$& 21.76/32 & $-0.28\pm 0.10$ & 24/32  \\
$\eta_{0}$ (Non-linear case) & $-0.28 \pm 0.24$& 21.2/32 & $-0.45 \pm 0.15$ & 23.7/32\\
\hline \hline
\end{tabular} \label{table3}
\end{center}
\end{table*}

Interestingly, we also obtained conflicting results between the
two light-curve fitters.  In our analysis involving the elliptical
De Filippis et al.\ (2005) sample (for both light-curve fitters), the
DD relation validity is obtained at 1$\sigma$ (c.l.) with the SALT2
method. However, it is only marginally compatible with MLC2K2 method
because the $\eta_0$ values are considerably negative. This result
points to an overestimated luminosity distances when SNe Ia are
analyzed with the MLC2K2 method. This result is not surprising since the
De Filippis et al.\ (2005) sample agrees with the DD
relation validity in the context of the $\Lambda$CDM model (WMAP7)
(see section \ref{sec:meth1}) and the SALT2 light-curve fitter also
favors the same cosmology according to the analyses of Kessler et
al.\ (2009) and Sollerman et al.\  (2009). Curiously,  the best
verification of the DD relation validity is obtained by confronting
the spherical sample of De Filippis et al.\ (2005) with SNe Ia when
the MLC2K2 light-curve fitter is adopted. This probably occurs
because both distances are overestimated compared to distances
from the elliptical model and the $\Lambda$CDM model (WMAP7). For the
Bonamente et al.\ (2006) sample the DD relation validity  is
marginally compatible (2$\sigma$) with SALT2 and at 3$\sigma$ with
MLCS2K2.

\section{Searching for the true geometry of clusters: a joint analysis}

In the previous sections we have discussed  how the validity of
the  DD relation  can be used to fix the true cluster geometry by
using two independent approaches: (i) The model-dependent test based
on the SZE/X-ray technique in the context of the $\Lambda$CDM model
(WMAP7 analysis), and (ii) The model-independent test using the
combination of SZE/X-ray for galaxy clusters and SNe Ia data. For
the former  case the elliptical description provides the best fit
supporting the standard DD relation in $\Lambda$CDM (Komatsu et al.\
2011). On the other hand, the isothermal spherical $\beta$ model is
more compatible with the validity of the DD relation in the
model-independent test involving supernovas. Since likelihoods for
both descriptions in both approaches (see figs 2a and 5a) are 
compatible with each other in at least  2-sigma, it is convenient to
perform  a joint analysis involving these independent and complementary
treatments in order to choose the better description.

Naturally, one may argue that the joint analysis will be plagued by
the underlying tension between SNe and CMB data. In this context, we
recall that this tension was recently discussed by taking into account
weak-lensing effects caused by inhomogeneities at low and intermediate
redshifts by Amendola et al.\ (2010. By using the Union data (2008),
they showed that the inclusion of lensing moves the best-fit model
significantly toward the flat LCDM  model (owing to corrections 
induced by the lens convergence on the distance modulus). This treatment
is beyond the scope of our paper. Nevertheless, this treatment is valid
in the same sense as the early analysis involving SNe Ia and CMB were to
the cosmic concordance model. As we shall see, the joint analysis below
confirms that the elliptical isothermal model provides the best description
for galaxy clusters. 

Now, by adding the $\chi^2$ statistics for the different approaches,
the  resulting likelihoods for the  $\eta_0$ parameter were again
obtained. The main findings  for both parametrizations (linear and
non-linear cases) are summarized in Table 4. Note that we restricted
our attention to the De Filippis et al.\ (2005) samples (elliptical
and spherical isothermal  assumptions).

From Table 4, we see that the elliptical geometry agrees better
with the strict validity of the duality relation when both approaches
are considered. This result remains valid regardless of
the supernova sample (Constitution or SDSS) adopted, the
parameterization (linear or non-linear), or the kind of light
fitter used.

\begin{table*}[ht!]
\caption{{The $\eta_0$ values for our joint analysis. All errors
are at $1\sigma$ and the $\chi^2/d.o.f.$ is given in parentheses}.}
\begin{center}
\begin{tabular} {|c|c|c|c|}
\hline\hline  Isothermal elliptical $\beta$ model & WMAP+SNe Ia &
WMAP+SNe Ia&WMAP+SNe Ia\\
 & Constitution&SDSS(SALT2)&SDSS(MLCs2k2)
\\ \hline \hline
$\eta_{0}$ (Linear case)& $-0.096^{+0.096}_{-0.084} (0.96)$& $-0.08^{+0.09}_{-0.09} (0.93)$& $-0.11^{+0.10}_{-0.12} (0.97)$ \\
$\eta_{0}$ (Non-linear case)& $-0.14^{+0.13}_{-0.13} (0.97)$&$-0.12^{+0.12}_{-0.13} (0.95)$ & $-0.177^{+0.143}_{-0.127} (0.96)$ \\
\hline\hline  Isothermal spherical $\beta$ model  & WMAP + SNe Ia & WMAP+SNe Ia&
WMAP+SNe Ia \\
& Constitution&SDSS(SALT2)&SDSS(MLCS2k2)
\\ \hline \hline
$\eta_{0}$ (Linear case) & $0.178^{+0.097}_{-0.108} (0.76)$ & $0.195^{+0.105}_{-0.115} (0.77)$& $0.16^{+0.12}_{-0.10} (0.75)$\\
$\eta_{0}$ (Non-linear case) & $0.263^{+0.157}_{-0.148} (0.75)$ & $0.29^{+0.14}_{-0.15} (0.76)$ & $0.24^{+0.152}_{-0.148} (0.75)$ \\
\hline\hline
\end{tabular} \label{table4}
\end{center}
\end{table*}

In Fig. 9, we plotted the likelihoods function for the $\eta_0$
parameter as obtained from our joint analysis by considering
again the De Filippis et al.\ (2005) samples (elliptical and
spherical isothermal geometries). Fig. 9a is the joint analysis for
the WMAP + SNe Ia  by using the Constitution sample while in Figs.
9b and 9c, we considered SDSS sample, but now taking into account the
different light-curve fitters  (SALT2 and MLCS2k2). The horizontal
lines are cuts in the regions of 68.3 (1$\sigma$) per cent and 95.4
(2$\sigma$) per cent probability. Our joint analysis shows that an
elliptical geometry is suggested by the existing data even when we
take into account the SNe Ia light-curve fitters. Because several SZE
surveys are in progress, our results call attention to a basic
difficulty involving the spherical assumption to describe the
cluster morphology when independent cosmological probes are
considered.

\section{Conclusions}
\label{sec:conclusions}

We explored some physical consequences of a deformed
distance-duality relation, $\eta(z) = D_{L}(1+z)^{-2}/D_{A}$, based
on observations of galaxy clusters using SZE and X-ray surface
brightness. The $\eta(z)$ parameter was described by two distinct
forms, $\eta = 1+\eta_{0}z$ and $\eta = 1+\eta_{0}z/(1+z)$, thereby
recovering the standard equality between distances for extremely low
redshifts.

Initially, we discussed the consistency between the strict validity
of the distance-duality relation and the underlying assumptions
about the  galaxy cluster geometries.  In our analysis we  used
angular diameter distances from galaxy clusters of the De Filippis
et al.\ (2005) and Bonamente et al.\ (2006) samples. The former
sample consists of 25 data sets where the galaxy clusters were described
by an isothermal elliptical  $\beta$ model and a spherical $\beta$
model, whereas the latter sample consists of 38 data sets where a
non-isothermal spherical double $\beta$ model was used. We showed
that the elliptical geometry is more consistent with no
violation of the distance duality relation in the context of
$\Lambda$CDM (WMAP7 data). In the elliptical case of the De Filippis
et al.\ (2005) sample (see Fig.\ 2a and table \ref{table1}) we found
$\eta_{0} = - 0.056^{+ 0.1}_{- 0.1}$ and  $\eta_{0} = -0.088^{+
0.14}_{- 0.14}$ for linear and non-linear parametrizations at
1$\sigma$ (statistical plus systematic errors), respectively. The
analysis with the isothermal spherical $\beta$ model led to higher
and positive $\eta_0$ values  than those from elliptical $\beta$
model, indicating that the estimated angular distances  with the
spherical $\beta$ model are overestimated compared to those
from $\Lambda$CDM (WMAP7). On the other hand, the non-isothermal
spherical double $\beta$ model (see Fig.\ 2b) was only marginally
compatible with $\eta_{0} =0$, such as $\eta_{0} = -0.12^{+ 0.12}_{-
0.12}$ and $\eta_{0} = -0.175^{+ 0.175}_{- 0.175}$ for linear and
non-linear parameterizations at 2$\sigma$ (statistical plus
systematic errors), respectively. In this way, our analysis revealed
that the elliptical model is compatible with the duality relation at
1$\sigma$, whereas the non-isothermal spherical model (in Bonamente
et al.\ 2006 sample) is only marginally compatible at 3$\sigma$ {in
the $\Lambda$CDM framework}.

Furthermore, we discussed a new and model-independent cosmological test
for the distance-duality relation. We considered the three
angular diameter distance samples from galaxy clusters, which are
obtained by using SZE and X-ray surface brightness, together the
luminosity distances given by two subsamples of SNe Ia taken from
the Constitution data.  For each subsample, the redshifts of the
SNe Ia  were carefully chosen to coincide with those of the
associated galaxy cluster sample ($\Delta z<0.005$). Our results
showed that the confrontation between the angular diameter distances
from the De Filippis et al.\ (2005) sample (elliptical model) with SNe
Ia data pointed to a moderate violation of the reciprocity relation
(the DD relation was marginally satisfied in $2\sigma$) and the
$\eta_0$ values were predominantly negative. Since $(1+z)^{2}D^{\:
data}_{A}(z)/D_{L}(z) \propto 1+ \eta_{0}z$, negative $\eta_0$
values indicated luminosity distances overestimated with relation to
angular diameter ones. This tension between Sne Ia and the
elliptical De Filippis et al.\ (2005) sample arises because SNe Ia
samples prefer universes with higher $\Omega_{\Lambda}$ values than
the WMAP7 results. When an isothermal spherical $\beta$ model
is adopted,  the DD relation had been satisfied at $1\sigma$
with $\eta_0$ values preferably positive. Seemingly, this concordance
occurred only because an isothermal spherical $\beta$ model yields
angular distances overestimated in comparison to an elliptical model
and WMAP7 results. For the Bonamente et al.\ (2006) sample, where a
non-isotermal spherical double $\beta$ model was assumed to describe
the clusters, we saw that the DD relation is not obeyed even at
3$\sigma$. For this case we obtained $\eta_{0} = -0.42^{+ 0.34}_{- 0.34}$
and $\eta_{0} = -0.66^{+ 0.5}_{- 0.5}$ for linear and non-linear
$\eta(z)$ parameterizations in 3$\sigma$ (statistical plus systematic
errors), respectively.

Moreover, by using the SNe Ia data from SDSS-II (Kessler et al.\
2009), we explored the influence of the two different SNe Ia
light-curve fitters in our model-independent cosmological test for
the DD relation. In this way, we considered two subsamples of SNe
Ia taken from SDSS-II (2007) for $D_L$ where, again, the
SNe Ia redshifts of each subsample were carefully chosen to
coincide with those of the associated galaxy cluster sample
($\Delta z<0.005$). In this case, each subsample of SNe Ia was
analyzed by the MLCS2K2 and SALT2 light-curve fitters separately.
Interestingly, we obtained conflicting results between the two
light-curve fitters. For both light-curve fitter-method the best
fits are obtain in an analysis involving the elliptical De Filippis et
al.\ (2005) sample. In this case, the DD relation validity was
obtained in 1$\sigma$ with the SALT2 method and it was marginally
compatible with MLC2K2 method with $\eta_0$ values considerably
negative. This result points to overestimated luminosity
distances when SNe Ia are analyzed with MLC2K2 method. For this
light-curve fitter, the isothermal spherical $\beta$ model  provided
that the DD relation was satisfied at $1\sigma$.  For all tests of
the Bonamente et al.\ (2006) sample, the DD relation validity is
obtained marginally at 2$\sigma$ (SALT2) and at 3$\sigma$ (MLCS2K2).
We stress that our results are independent of the redshifts chosen
to realize the test (of the galaxy clusters or SNe Ia), thereby
showing the robustness of the method and supporting $\Delta z < 0.005$
as a fair choice.

Finally, we applied a joint analysis involving these
independent and complementary treatments by adding the $\chi^2$
statistics for the different approaches. Interestingly, we saw that
the elliptical geometry agrees better with the strict
validity of the duality relation when both approaches were
considered. This result remains valid regardless of the adopted
SNe Ia sample (Constitution or SDSS), the parameterization (linear
or non-linear), or the kind of light-curve  fitter used (see
Table 4). The non-isothermal spherical double $\beta$ model was only
marginally compatible with the DD relation validity in all
treatments.

Summarizing, the statistical analysis presented here provides new
evidence that the true geometry of clusters has an elliptical form.
In principle,  it is remarkable that a local property such as the
geometry of galaxy clusters may be constrained by a global
argument like the one provided by the cosmological distance-duality
relation. This result also reinforces the interest in the
observational search for SZE and X-ray from clusters at high
redshifts. In the near future, when more and larger samples with
smaller statistical and systematic uncertainties become available,
the method proposed here, based on the validity of the
distance-duality relation, can improve the limits on the possible
cluster geometries and explore systematic errors in SNe Ia and
galaxy clusters observations.
\newline

\centerline{ Acknowledgments} RFLH is supported by FAPESP (No.\
07/52912-2) and JASL is partially supported by CNPq (No.\
306054/2010-  and FAPESP (No. 04/13668-0).

\begin{figure*}[ht!]
   \centering
       \includegraphics[width=0.45\linewidth]{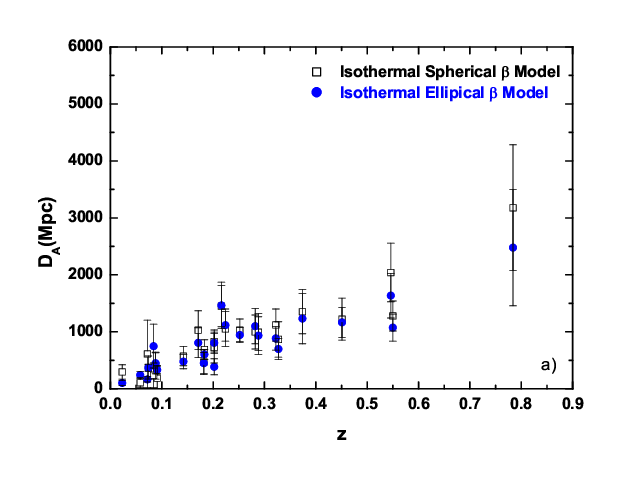}
       \includegraphics[width=0.45\linewidth]{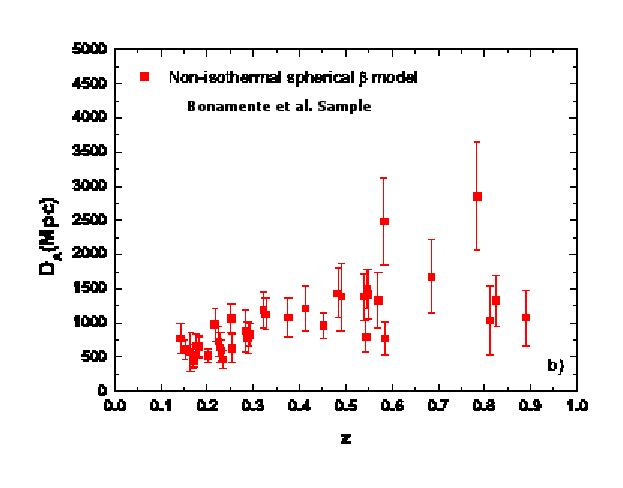}
   \caption{a) Filled (blue) circles and open (black) squares with the
associated error bars (only statistical errors) stand for the De Filippis
et al.\ (2005) samples: isothermal elliptical $\beta$ model and  isothermal
spherical $\beta$ model, respectively. b) The sample of Bonamente et al.\
(2006) where a non-isothermal spherical $\beta$ model was used to describe
the galaxy clusters.}
\end{figure*}
\begin{figure*}[ht!]
   \centering
       \includegraphics[width=0.45\linewidth]{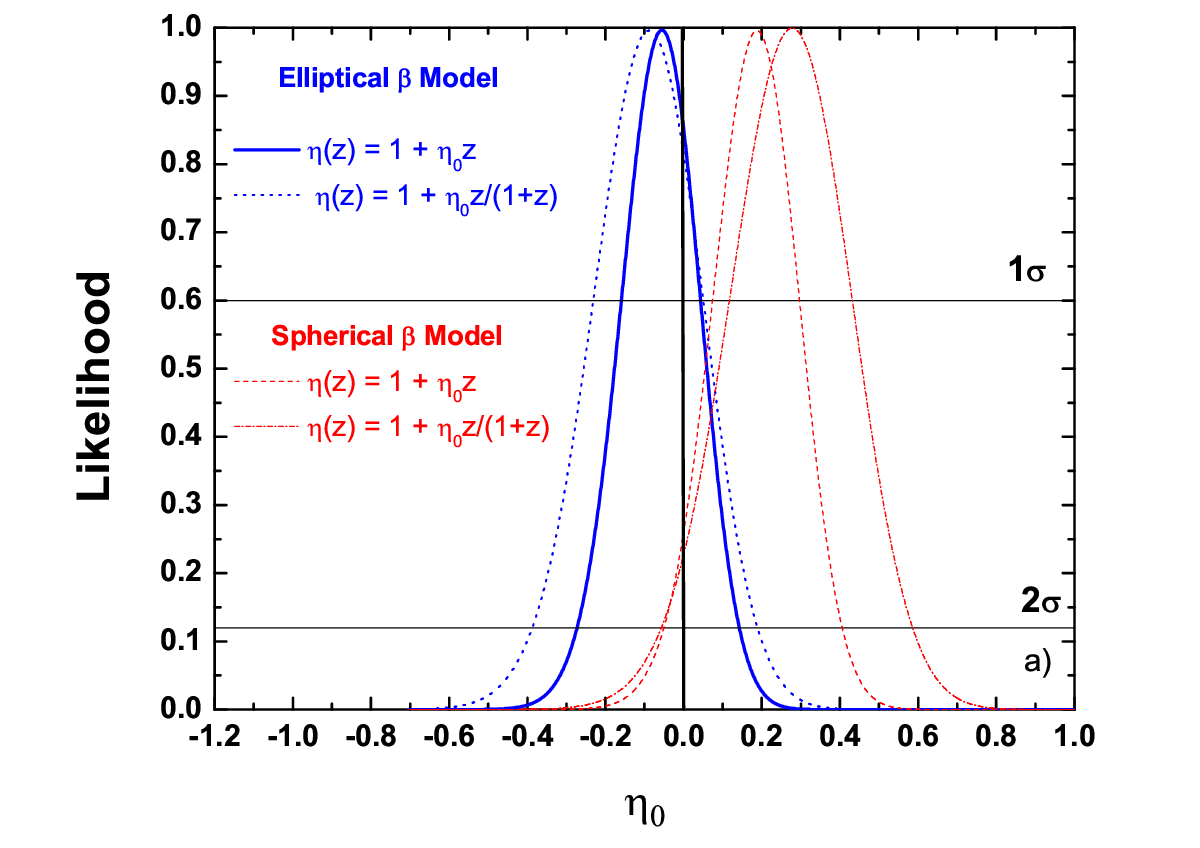}
       \includegraphics[width=0.40\linewidth]{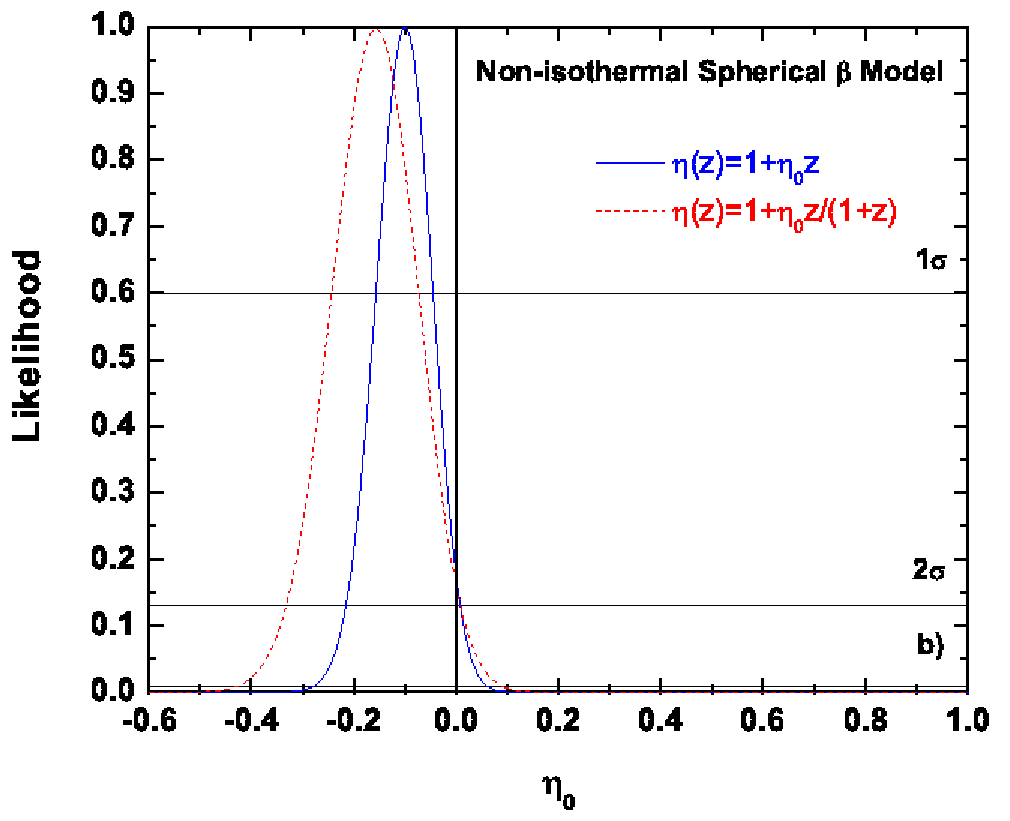}
   \caption{ a) Likelihood distribution functions for
De Filippis  et al.\ (2005) sample. The solid  and the dotted blue
lines are likelihood functions corresponding to linear and
non-linear $\eta(z)$ parametrizations for the isothermal elliptical
$\beta$ model. The dashed and the dashed-dotted red lines are
likelihood functions corresponding to linear and non-linear
parametrizations for the isothermal spherical $\beta$ model. b) The
likelihood distribution functions for  Bonamente et al.\ (2006)
sample. The solid blue line  and the dotted red line correspond to
linear and non-linear parametrizations. }
\end{figure*}

\begin{figure*}[tb!]
   \centering
       \includegraphics[width=0.45\linewidth]{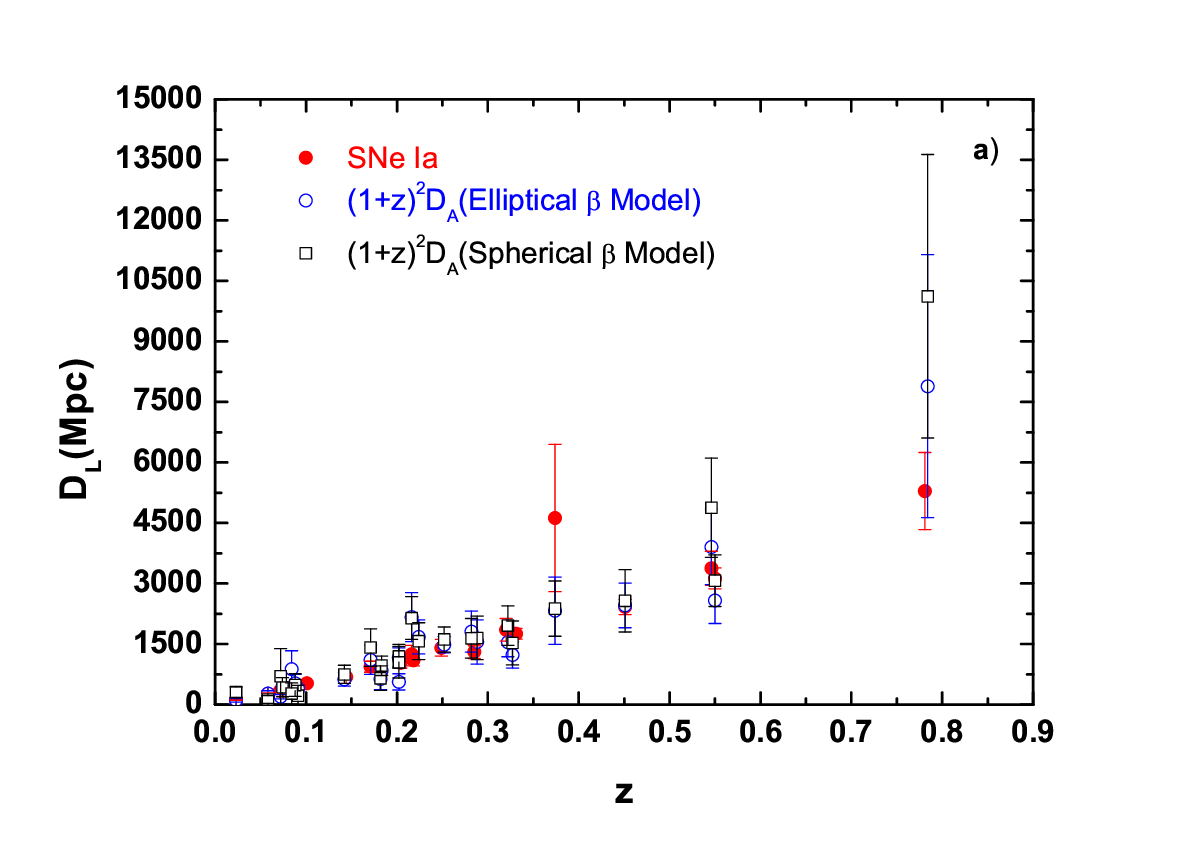}
       \includegraphics[width=0.45\linewidth]{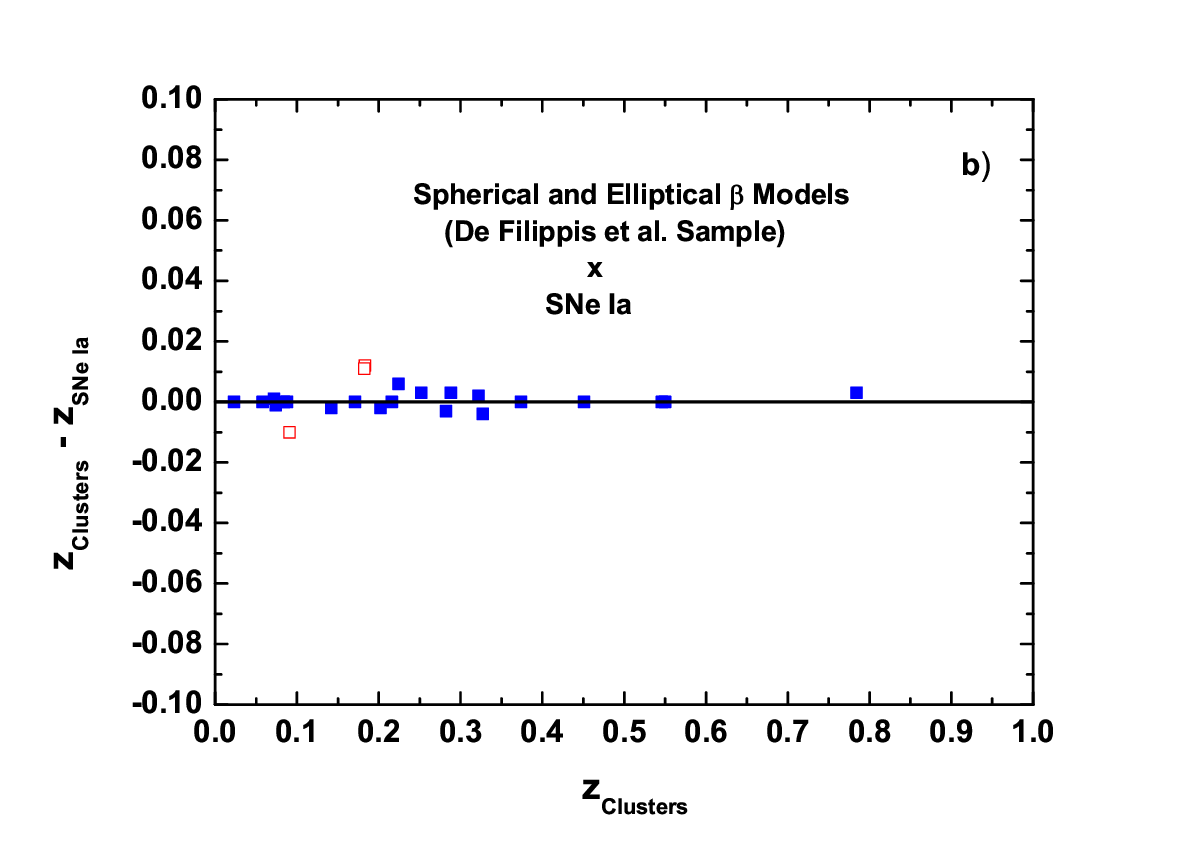}
   \caption{a) Galaxy clusters and SNe Ia data. The open blue, open
black, and filled red circles with the associated error bars stand for
the galaxy clusters described with elliptical and spherical $\beta$
models of the De Filippis et al.\ (2005) (statistical + systematical
errors) and SNe Ia (only statistical errors) samples, respectively.
b) The redshift subtraction for the same pair of cluster-SNe Ia samples.
The open squares represent the pairs of points for which
$\Delta z \approx 0.01$.}
\end{figure*}
\begin{figure*}[tb!]
   \centering
       \includegraphics[width=0.45\linewidth]{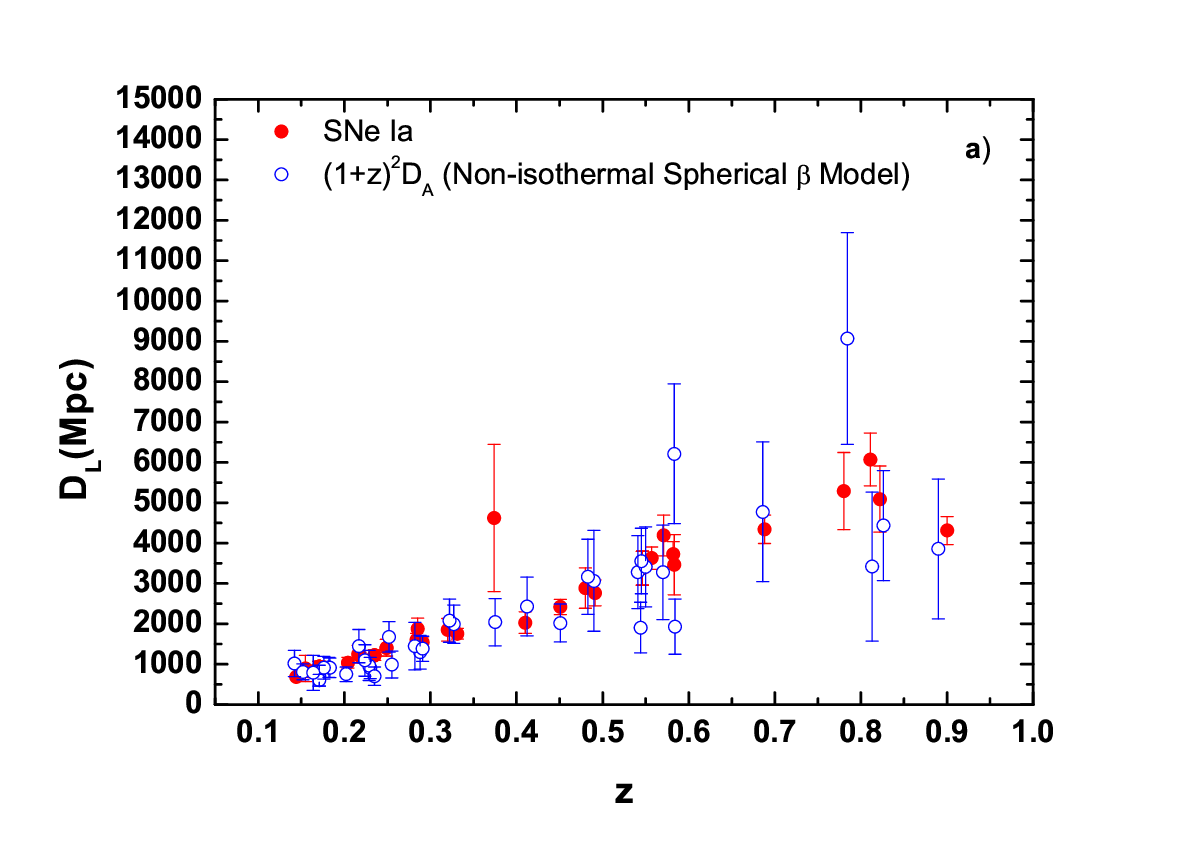}
       \includegraphics[width=0.45\linewidth]{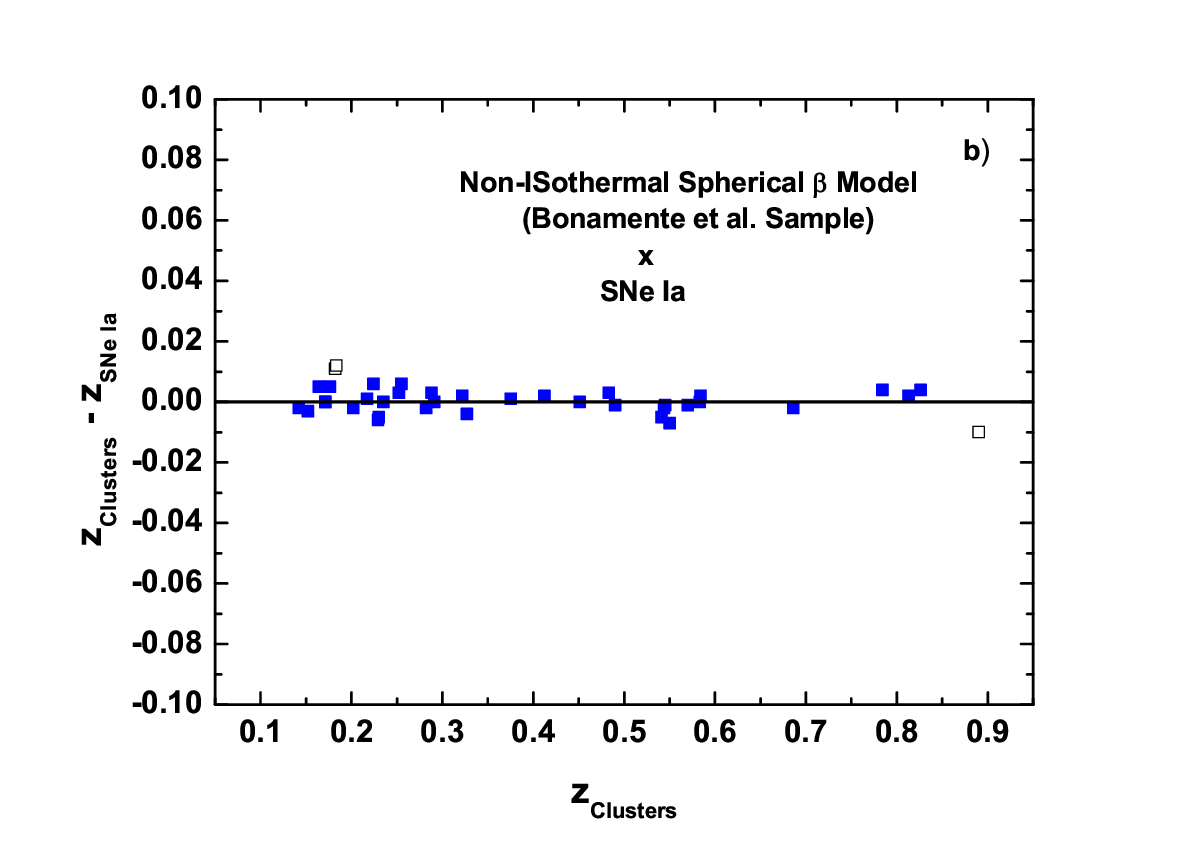}
   \caption{a) Galaxy clusters and SNe Ia data. The open (blue) and
filled (red) circles with the associated error bars stand for the
Bonamente et al.\ (2006) (statistical + systematical errors) and SNe
Ia (only statistical errors) samples, respectively. b) The redshift
subtraction for the same pair of cluster-SNe Ia samples. As in Fig.\
1b, the open squares represent the pairs of points with the biggest
difference in redshifts ($\Delta z \approx 0.01$).}
\end{figure*}
\begin{figure*}[tb!]
{\includegraphics[width=85mm, angle=0]{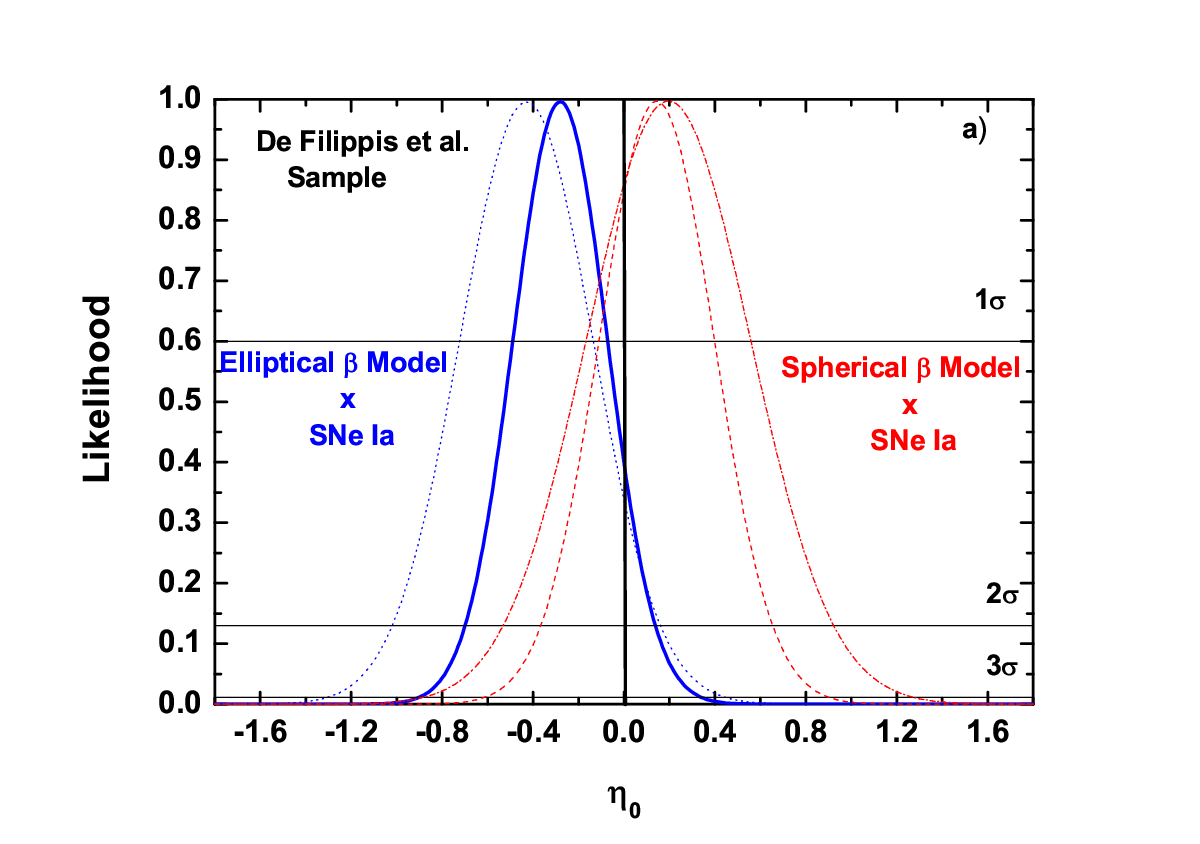}
\includegraphics[width=77mm, angle=0]{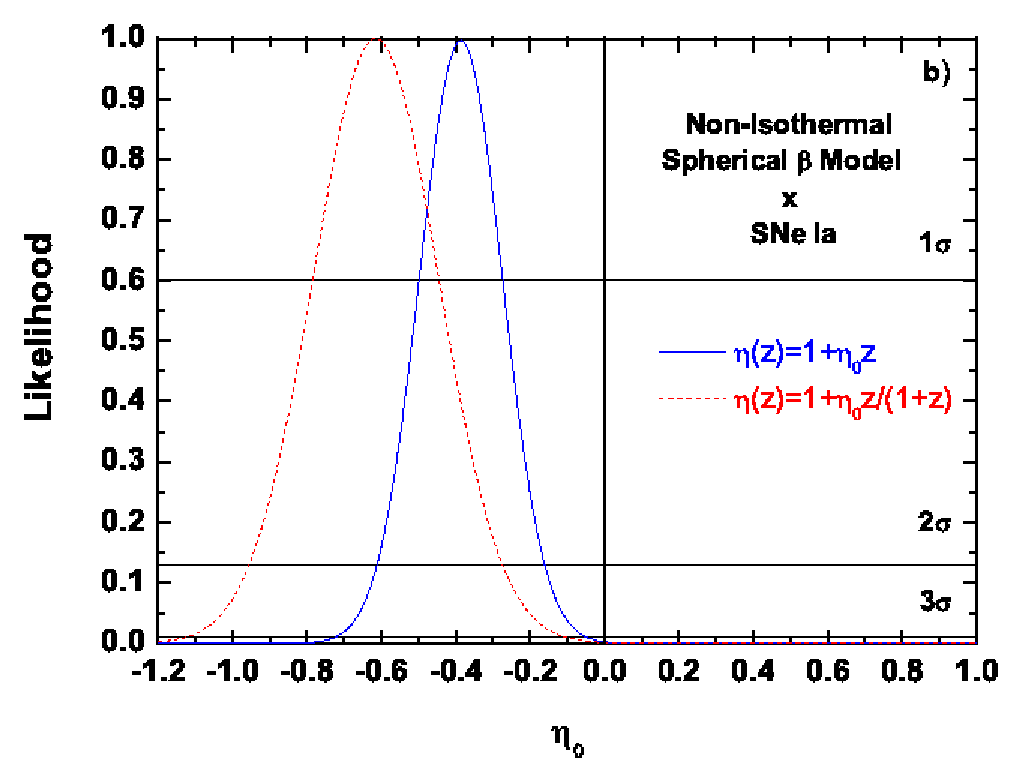}
\hskip 0.1in} \caption{ {{a)}} Likelihood distribution functions
for De Filippis  et al.\ (2005) sample. The solid  and the dotted
blue lines are likelihood functions for linear and non-linear
parametrizations corresponding to the isothermal elliptical $\beta$
model. The dashed and the dashed-dotted red lines are likelihood
functions for linear and non-linear parametrizations corresponding
to the isothermal spherical $\beta$ model. {{b)}} The likelihood
distribution functions for  Bonamente et al.\ (2006) sample.}
\label{fig:Analysis2}
\end{figure*}

\begin{figure*}[tb!]
   \centering
       \includegraphics[width=0.45\linewidth]{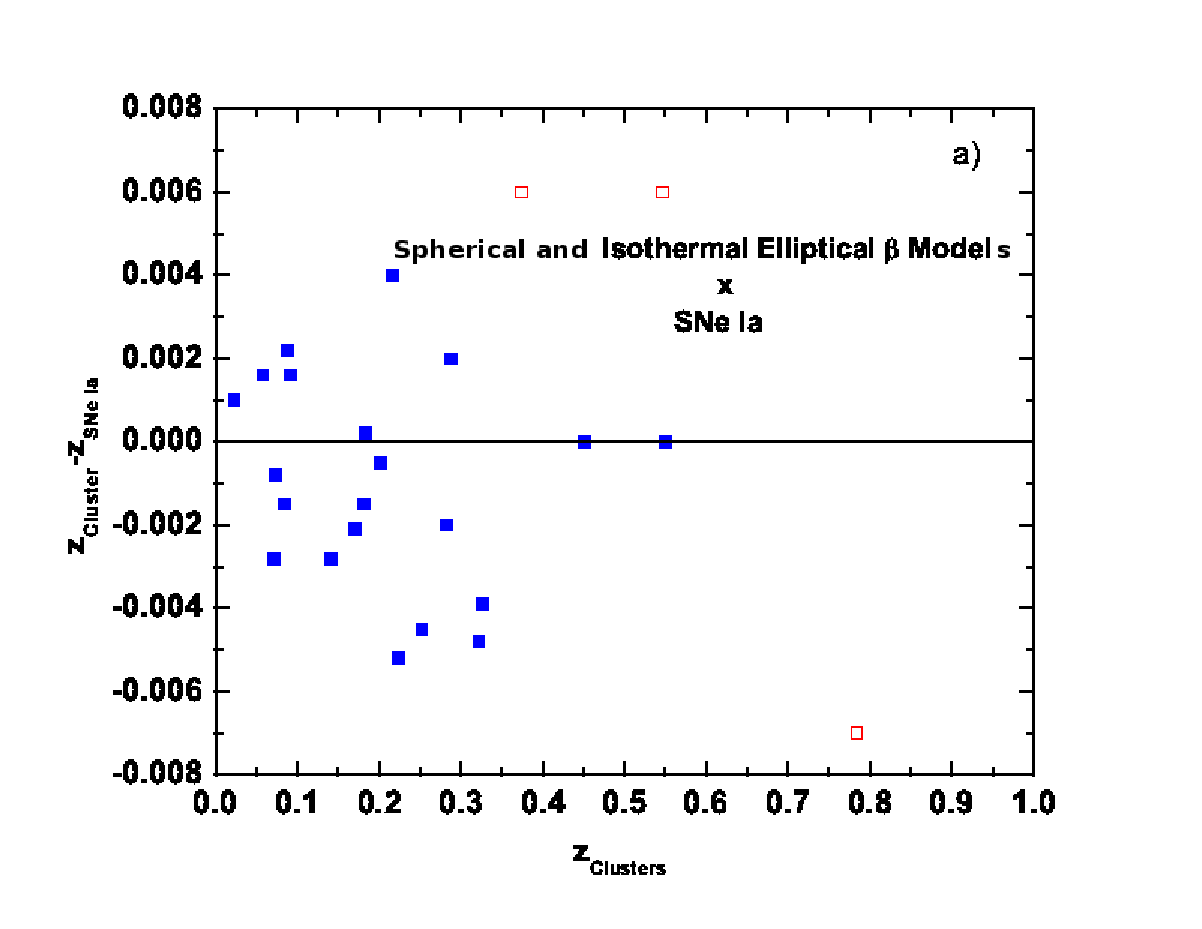}
       \includegraphics[width=0.45\linewidth]{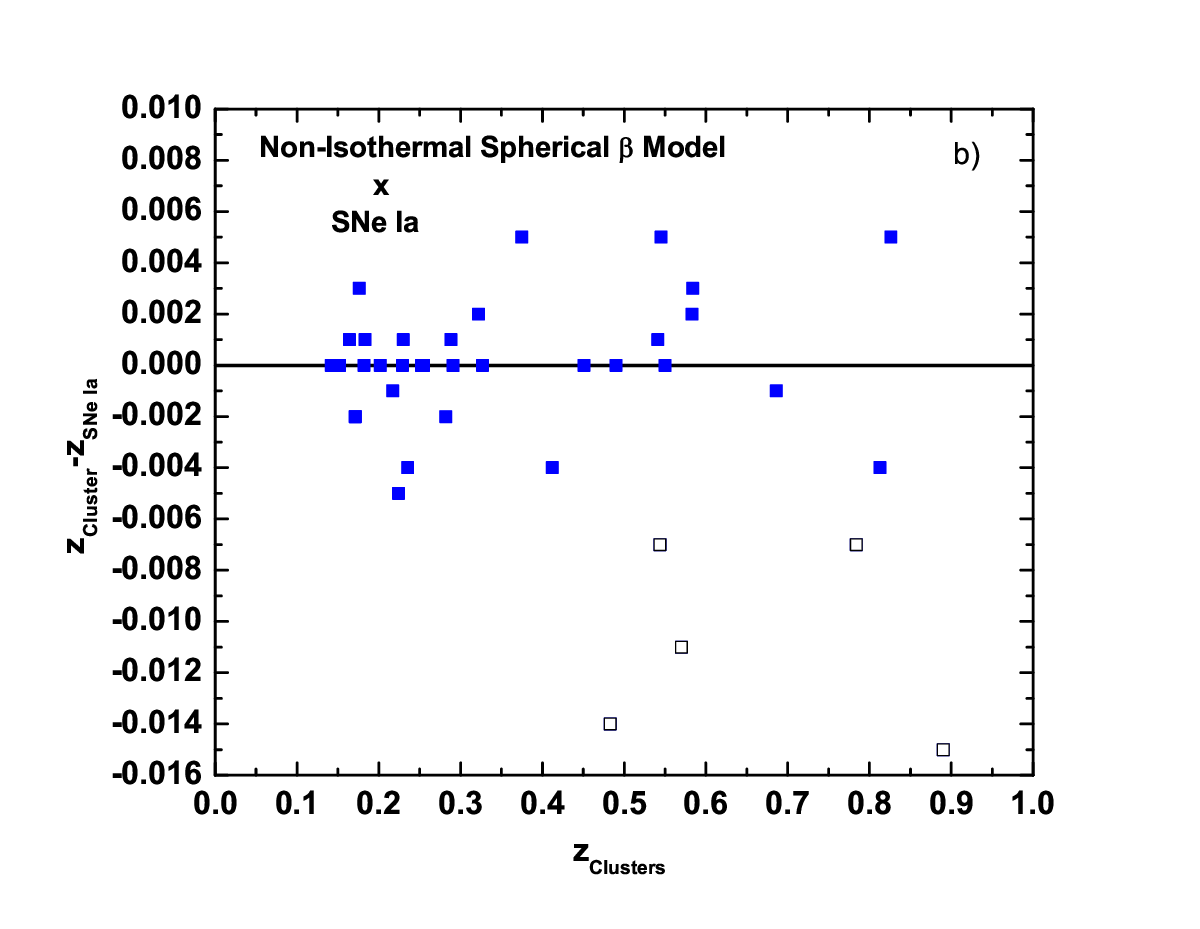}
   \caption{a) Redshift subtraction for the same pair of cluster-SNe
Ia samples of De Filippis et al.\ (2005). The open squares represent
the pairs of points for which $\Delta z > 0.005$. b) The redshift
subtraction for the same pair of cluster-SNe Ia samples for Bonamente
et al.\ (2006). The open squares represent the pairs of points for
which $\Delta z > 0.005$.}
\end{figure*}

\begin{figure*}[tb!]
  \centering
      \includegraphics[width=0.45\linewidth]{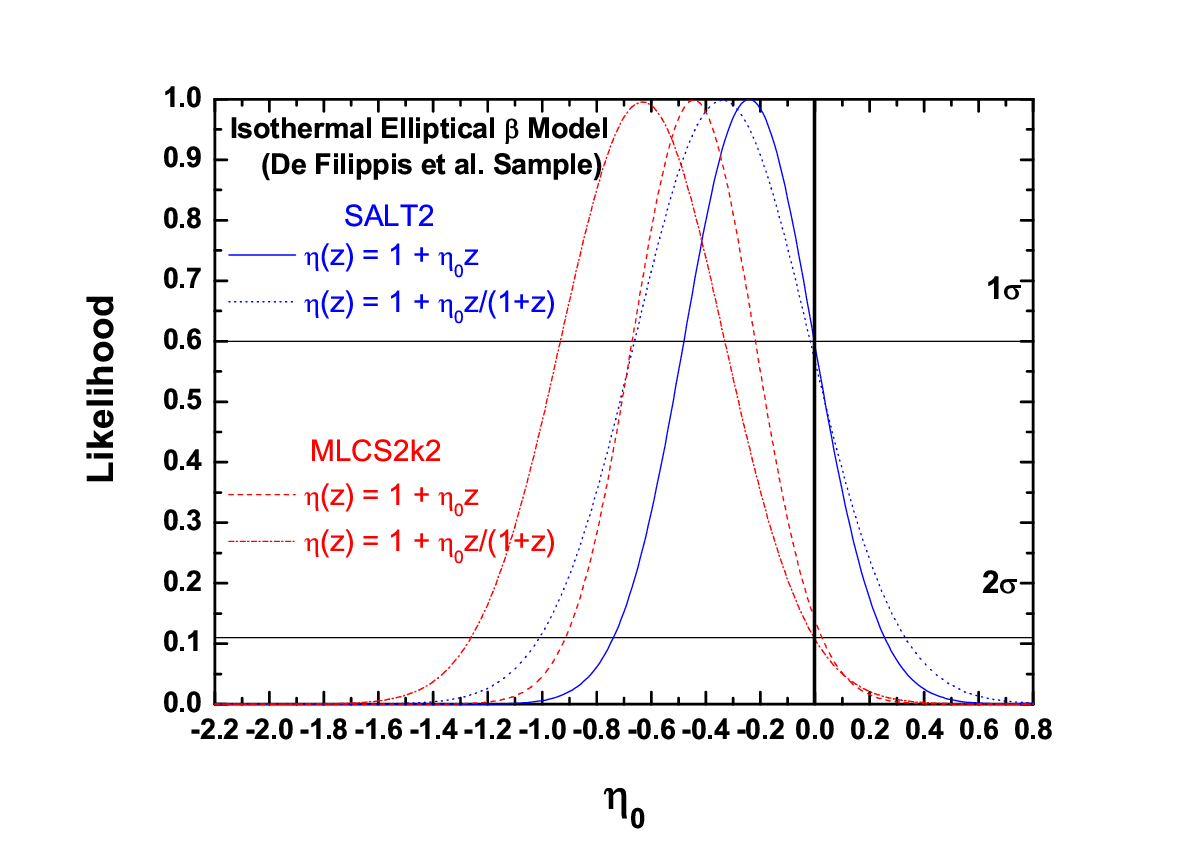}
      \includegraphics[width=0.45\linewidth]{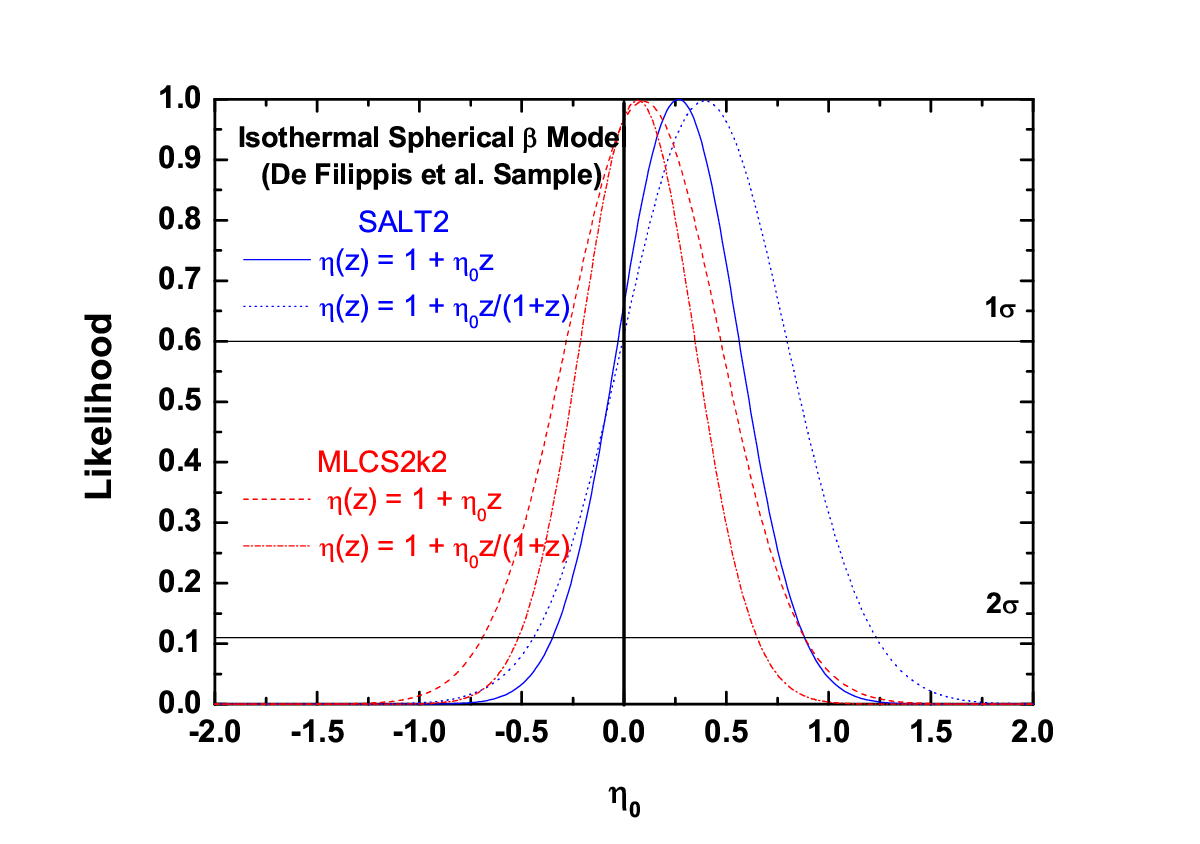}
  \caption{{{a)}} Likelihood distribution functions for the De
Filippis et al.\ (2005) sample (elliptical $\beta$ model) for
both $\eta(z)$ parametrizations and both SNe Ia light-curve fitters.
{{b)}}  The likelihood distribution functions for the De Filippis et
al.\ (2005) sample (spherical $\beta$ model) for both $\eta(z)$
parametrizations and both SNe Ia light-curve fitters.}
\end{figure*}

\begin{figure*}[tb!]
  \centering

      \includegraphics[width=0.40\linewidth]{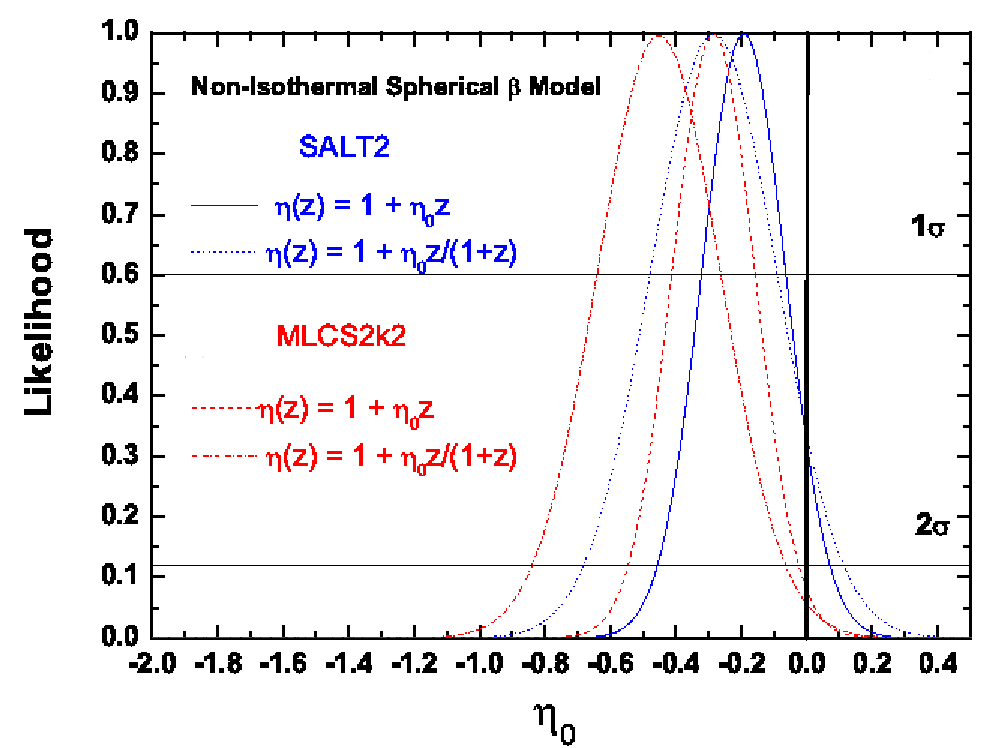}
  \caption{Likelihood distribution functions for the
Bonamente et al.\ (2006) sample for both parametrizations and both
SNe Ia light-curve fitters.} 
\end{figure*}

\begin{figure*}[tb!]
  \centering
      \includegraphics[width=0.45\linewidth]{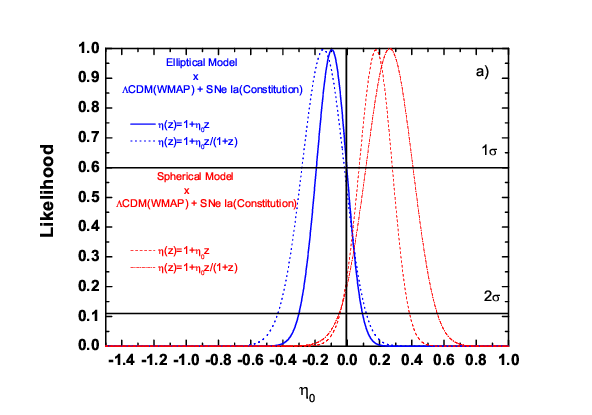}
      \includegraphics[width=0.45\linewidth]{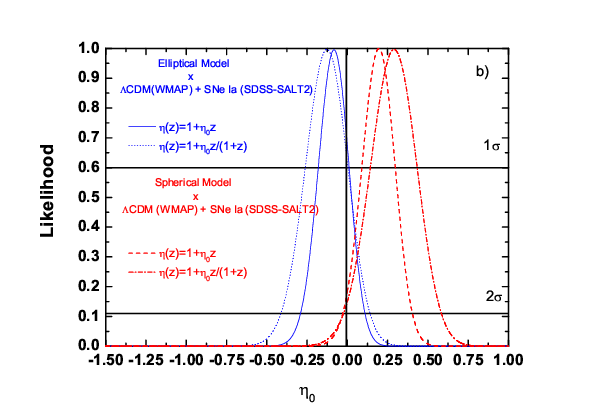}
       \includegraphics[width=0.45\linewidth]{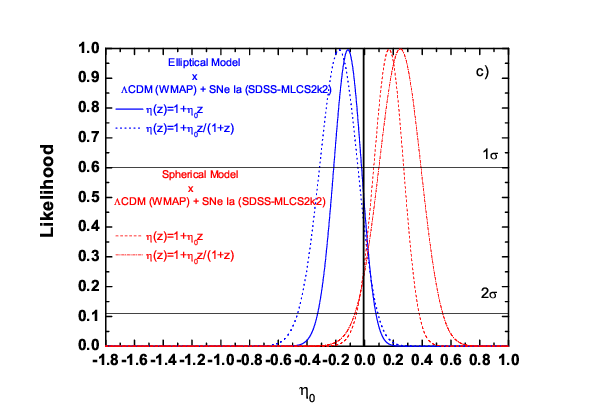}
  \caption{{{ a)} Joint analysis for the De Filippis et al.\ (2005)
samples and WMAP + SNe Ia (Constitution Sample). {b)} Joint analysis for
the De Filippis et al.\ (2005) samples and WMAP + SNe Ia (SDSS-SALT2).
{c)} Joint analysis for the De Filippis et al.\ (2005) samples and
WMAP + SNe Ia (SDSS-MLCS2k2).}}

\end{figure*}


\begin{thebibliography}{99}
\bibitem{albani} Albani, V. V. L., Iribarrem, A. S., Ribeiro, M. B., \&
        Stoeger, W. R.\ 2007, ApJ, 657, 760 [astro-ph/0611032]
\bibitem{union2}Amanullah, R., et al.\ 2010, ApJ, 716, 712
\bibitem{Astier} Astier, P., et al.\ 2006, A\& A, 447, 31
\bibitem{lverde} Avgoustidis, A., Burrage, C., Redondo, J., Verde, L.,
        \& Jimenez, R.\ [astro-ph/10042053]
\bibitem{bk04} Basset, B. A., Kunz M.\ 2004, Phys.\ Rev.\ D, 69, 101305
        [astro-ph/0312443v2]
\bibitem{ben}Bengochea, G. R. 2011, PLB, 696,5
\bibitem{Boname06} Bonamente, M., et al.\ 2006, ApJ, 647, 25
\bibitem{Boname04} Bonamente, M., et al.\ 2004, ApJ, 614, 56
\bibitem{caval} Cavaliere, A., \& Fusco-Fermiano, R.\ 1978, A\&A., 667, 70
\bibitem{CMS07} Cunha, J. V., Marassi, L., Santos, R. C.\ 2007, IJMPD, 16,
        403 [astro-ph/0608686]
\bibitem{CML07}Cunha, J. V., Marassi, L., \& Lima, J.A.S.\ 2007, MNRAS,
        379, L1 [astro-ph/0611934]
\bibitem{daly} Daly, R. A., \& Djorgovski, S. G.\ 2003, ApJ, 597, 9
\bibitem{Davis}Davis, T. M., et al.\ 2007, ApJ, 666, 716
\bibitem{bem06} De Bernardis, F., Giusarma, E., \& Melchiorri, A.\ 2006,
        IJMPD, 15, 759 [gr-qc/0606029v1]
\bibitem{dagostini} D'Agostini, G.\ 2004 [physics/0403086]
\bibitem{DeFilippis05}  De Filippis, E., Sereno, M., Bautz, M.W., \&
        Longo, G.\ 2005, ApJ, 625, 108
\bibitem{ebelin} Ebeling, H., et al.\ 1996, MNRAS, 281, 799
\bibitem{ellis71} Ellis G. F. R.\ 1971, ``Relativistic Cosmology'',
        Proc.\ Int.\ School Phys.\ Enrico Fermi, R. K.\ Sachs (ed.),
        pp. 104-182 (Academic Press: New York); reprinted in  Gen.\
    Rel.\ Grav., 41, 581, 2009
\bibitem{ellis07} Ellis, G. F. R.\ 2007, Gen.\ Rel¿\ Grav., 39, 1047
\bibitem{eth33} Etherington, I. M. H.\ 1933, Phil.\ Mag., 15, 761;
        reprinted in Gen.\ Rel.\ Grav., 39, 1055, 2007
\bibitem{Goncalves2011} Gol\c{c}alves, R. S., Holanda, R. F. L.  \& Alcaniz, J. S. 2011 [arXiv:1109.2790]
\bibitem{G94} Gurvitz, L. I.\ 1994, ApJ, 425, 442
\bibitem{G99} Gurvitz, L. I., Kellerman, K.I., \& Frey, S.\ 1999, A\&A,
        342, 378
\bibitem{guy}{Guy, J., et al.\ 2007 A \& A, 466, 11}
\bibitem{jha} {Jha, S., Riess, A. G., \& Kirshner, R. P. 2007, ApJ, 659, 122}
\bibitem{Hicken}{Hicken, M., et al.\ 2009, ApJ, 700, 1097}
\bibitem{Holandaa}Holanda, R.~F.~L., Lima, J.~A.~S., \& Ribeiro, M.~B.\
       2010, ApJL, 722, L233 [arXiv:1005.4458]
\bibitem{Holandab}Holanda, R.~F.~L., Lima, J.~A.~S., \& Ribeiro, M.~B.\
       2011, A\&A Letters, 528, L14 [arXiv:1003.5906]
\bibitem{itoh1998} Itoh, N., Kohyama, Y., \& Nozawa, S. 1998, ApJ, 502, 7-15
\bibitem{kessler} {Kessler, R., et al.\ 2009, ApJ, 185, 32 }
\bibitem{komatsu} Komatsu, E.,\ et al.\ 2011, ApJS, 192, 18
        (WMAP collaboration)
\bibitem{union} Kowalski, M., et al.\ 2008, ApJ, 749, 686
\bibitem{lunion} La Roque, S. J., et al.\ 2006, 652, 917
\bibitem{Li2011}Li, Z., Wu, P., \& Yu, H. 2011, 729, L14
\bibitem{Lian} Liang, N., Cao, S. \& Zhu, Z. 2011 [arXiv:1104.2497v1]
\bibitem{LA00} Lima, J. A. S., \& Alcaniz, J. S.\ 2000, A\&A, 357, 393
        [astro-ph/0003189]
\bibitem{LA02} Lima, J. A. S., \& Alcaniz, J. S.\ 2002, ApJ, 566, 15
        [astro-ph/0109047]
\bibitem{LAC03} Lima, J. A. S., Cunha, J. V., \& Alcaniz, J. S. 2003,
        Phys.\ Rev.\ D, 68, 023510 [astro-ph/0303388]
\bibitem{Lima2011} Lima, J. A. S., Cunha, J. V. \& Zanchin, V. T. 2011, ApJL accepted [arXiv:1110.5065]       
\bibitem{linder04} Linder, E.\ V.\ 2003, PRL, 90, 091301
\bibitem{Mantz09} Mantz, A., Allen, S. W., Rapetti, D., \& Ebeling, H.\
        2010, MNRAS, 406, 1759
\bibitem{Mas01} Mason, B. S., et al.\ 2001, ApJ, 555, L11
\bibitem{Meng} Meng, X-L., Zhang, T-J.,  \& Zhan, H. 2011 [arXiv:1104.2833v1]
\bibitem{Mohr} {Mohr}, J. J., {Mathiesen}, B \& {Evrard}, A. E. 1999, ApJ, 517, 627
\bibitem{morandi} Morandi, A., Pedersen, K., \& Limousin, M.\ 2010,
        ApJ, 713, 491
\bibitem{djain} Nair, R., Jhingan, S., \&  Jain, D.\ 2011 [arXiv:1102.1065]
\bibitem{navarro} Navarro, J. F., Frenk, C. S., \& White, S. D. M.\
        1997, 490, 493
\bibitem{Nozawa} Nozawa, S., Itoh, N., \& Kohyama, Y. 1998, ApJ, 508, 17-24
\bibitem{pad04} Padmanabhan, T., \& Choudury, R.\ 2003, MNRAS, 344, 823
\bibitem{percival} Percival, W., et al.\ 2010, MNRAS, 401, 2148
\bibitem{juracy} Rangel Lemos, L.J., \& Ribeiro, M. B.\ 2008, A\&A,
        488, 55 [arXiv:0805.3336]
\bibitem{Reese02} Reese, E. D., et al.\ 2002, ApJ, 581, 53
\bibitem{Reese04} Reese, E. D.\ 2004, in ``Measuring and Modeling the
        Universe'', ed.\ W. L.\ Freedman (CUP) p. 138 [astro-ph/0306073]
\bibitem{rb02} Reiprich, T. H., \& Bohringer, H.\ 2002, ApJL, 567, 716
\bibitem{rib92} Ribeiro, M.\ B.\ 1992, ApJ, 388, 1 [arXiv:0807.0866]
\bibitem{rib93} Ribeiro, M.\ B.\ 1993, ApJ, 415, 469 [arXiv:0807.1021]
\bibitem{rib05} Ribeiro, M.\ B.\ 2005, A\&A, 429, 65 [astro-ph/0408316]
\bibitem{RS03} Ribeiro, M. B., \& Stoeger, W.R.\ 2003, ApJ, 592, 1
       [astro-ph/0304094]
\bibitem{riess} Riess, A., et al.\ 2009, ApJ, 116, 1009
\bibitem{SL08} Santos, R.\ C., \& Lima, J.\ A.\ S.\ 2008, Phys.\
        Rev.\ D 77, 083505 [arXiv:0803.1865]
\bibitem{sarazin} Sarazin, C. L. 1988 in ``X-ray emission from clusters of galaxies''
   Cambridge Astrophysics Series, Cambridge University Press
\bibitem{SEF99} Schneider, P., Ehlers, J. \& Falco, E.E., Gravitational
        Lenses (Springer-Verlag, Berlin, 1992)
\bibitem{SHL09} Shang, C., Haiman, Z., \& Verde, L.\ 2009, MNRAS, 400,
        2, 1085 [arXiV:0908.2012v1]
\bibitem{SAL07} Silva, R., Alcaniz, J. S., \& Lima, J.A.S.\ 2007, IJMPD,
        16, 469
\bibitem{Sollerman} {Sollerman, J., et al.\ 2009, ApJ, 703, 1374}
\bibitem{spergel03} Spergel, D. N., et al.\ 2003, ApJS, 148, 175
\bibitem{lverde2} Stern, D., Jimenez, R., Verde, L., Kamionkowski, M.,
        \& Stanford, S.A., [arXiv:0907.3149]
\bibitem{SunZel72} Sunyaev, R. A., \& Zel'dovich, Ya.B. 1972, Comments
        Astrophys.\ Space Phys., 4, 173
\bibitem{uzan} Uzan, J. P., Aghanim, N., \& Mellier, Y. 2004, Phys.\
        Rev.\ D, 70, 083533 [astro-ph/0405620v1]
\bibitem{wei} Wei, H. 2010, PLB, 687, 286
\bibitem{wang}{Wang, J., \& White, S. D. M.\ 2009, MNRAS, 396, 709 }
\end{thebibliography}
\end{document}